\documentclass[a4paper,11pt]{article}
\pdfoutput=1 
\usepackage{siunitx}
\usepackage{jinstpub} 
                     
\usepackage{natbib}
\usepackage{lineno}

\usepackage{hyperref} 

\usepackage{subcaption}
\usepackage{xspace}

\def\geant      {\mbox{\textsc{Geant4}}\xspace}
\newcommand{\gev}{\ensuremath{\mathrm{\,Ge\kern -0.1em V}}\xspace}
\newcommand{\mev}{\ensuremath{\mathrm{\,Me\kern -0.1em V}}\xspace}
\newcommand{\kev}{\ensuremath{\mathrm{\,ke\kern -0.1em V}}\xspace}
\newcommand{\ev}{\ensuremath{\mathrm{\,e\kern -0.1em V}}\xspace}
\newcommand{\gevc}{\ensuremath{{\mathrm{\,Ge\kern -0.1em V\!/}c}}\xspace}
\newcommand{\mevc}{\ensuremath{{\mathrm{\,Me\kern -0.1em V\!/}c}}\xspace}
\newcommand{\gevcc}{\ensuremath{{\mathrm{\,Ge\kern -0.1em V\!/}c^2}}\xspace}
\newcommand{\mevcc}{\ensuremath{{\mathrm{\,Me\kern -0.1em V\!/}c^2}}\xspace}

\newcommand{\Nhits}{\ensuremath{N_{\mathrm{hits}}}\xspace}
\newcommand{\Nclus}{\ensuremath{N_{\mathrm{clus}}}\xspace}

\newcommand{\Nch}{\ensuremath{N_{\mathrm{ch}}}\xspace}

\newcommand{\tint}{\ensuremath{t_{\mathrm{int}}}\xspace}

\newcommand{\eg}{\mbox{\itshape e.g.}\xspace}
\newcommand{\ie}{\mbox{\itshape i.e.}\xspace}

\def\focal      {\mbox{FoCal}\xspace}

\def\epical      {\mbox{\textsc{Epical-2}}\xspace}

\def\mimosa      {\mbox{\textsc{Epical-1}}\xspace}

\newcommand{\Allpixtwo}{\ensuremath{\mathrm{Allpix^2}}\xspace}

\title{Performance of the Electromagnetic Pixel Calorimeter Prototype \epical}

\author[a]{J.~Alme,}
\author[b]{R.~Barthel,}
\author[b]{A.~van~Bochove,}
\author[c,d]{V.~Borshchov,}
\author[e]{R.~Bosley,}
\author[b]{A.~van~den~Brink,}
\author[b]{E.~Broeils,}
\author[f]{H.~B\"usching,}
\author[a]{V.N.~Eikeland,}
\author[a]{O.S.~Groettvik,}
\author[g]{Y.H.~Han,}
\author[b,h]{N.~van~der~Kolk,}
\author[g]{J.H.~Kim,}
\author[g]{T.J.~Kim,}
\author[g]{Y.~Kwon,}
\author[i]{M.~Mager,}
\author[j]{Q.W.~Malik,}
\author[b]{E.~Okkinga,}
\author[g]{T.Y.~Park,}
\author[b,1]{T.~Peitzmann,\note{Corresponding author.}}
\author[f]{F.~Pliquett,}
\author[c,d]{M.~Protsenko,}
\author[i]{F.~Reidt,}
\author[b]{S.~van~Rijk,}
\author[j]{K.~R{\o}ed,}
\author[f]{T.S.~Rogoschinski,}
\author[a]{D.~R\"ohrich,}
\author[b]{M.J.~Rossewij,}
\author[b]{G.B.~Ruis,}
\author[a,j]{E.H.~Solheim,}
\author[c,d]{I.~Tymchuk,}
\author[a]{K.~Ullaland,}
\author[e]{N.K.~Watson,}
\author[b,h]{H.~Yokoyama}

\affiliation[a]{Department of Physics and Technology, University of Bergen, Bergen, Norway} 
\affiliation[b]{Institute for Gravitational and Subatomic Physics (GRASP), Utrecht University/Nikhef, Utrecht, Netherlands}
\affiliation[c]{Research and Production Enterprise ``LTU'' (RPE LTU), Kharkiv, Ukraine}
\affiliation[d]{Bogolyubov Institute for Theoretical Physics, Kyiv, Ukraine}
\affiliation[e]{School of Physics and Astronomy, University of Birmingham, Birmingham, United Kingdom}
\affiliation[f]{Institut f\"ur Kernphysik, Johann Wolfgang Goethe-Universität Frankfurt, Frankfurt, Germany}
\affiliation[g]{Yonsei University, Seoul, Republic of Korea}
\affiliation[h]{Nikhef, National Institute for Subatomic Physics, Amsterdam, Netherlands}
\affiliation[i]{European Organization for Nuclear Research (CERN), Geneva, Switzerland}
\affiliation[j]{Department of Physics, University of Oslo, Oslo, Norway}
\emailAdd{t.peitzmann@uu.nl}

\abstract{The first evaluation of an ultra-high granularity digital electromagnetic calorimeter prototype using  1.0--5.8\gevc  electrons is presented.
The  $25\times10^6$ pixel detector consists of 24 layers of ALPIDE CMOS MAPS sensors, with a pitch of around \SI{30}{\micro\metre}, and has a depth of almost 20 radiation lengths of tungsten absorber. Ultra-thin cables allow for a very compact design.

The properties that are critical for physics studies are measured:
electromagnetic shower response, energy resolution and linearity.
The stochastic  energy resolution is comparable with the state-of-the art resolution for a Si-W calorimeter, with data described well by a simulation model using \geant and \Allpixtwo.
The performance achieved makes this technology a good candidate for use in the ALICE FoCal upgrade, and in general demonstrates the strong potential for future applications in high-energy physics.}


\arxivnumber{2209.02511}

\begin{document}
\maketitle
\flushbottom

\section{Introduction}
\label{sec:Introduction}
The fundamental principle underlying a digital electromagnetic calorimeter (DECAL) is that energy is measured by counting the number of charged particles in an electromagnetic shower. This reduces a source of uncertainty due to intrinsic fluctuations in the energy deposited, which can be significant in conventional calorimeters.
In a DECAL, the charged particle multiplicity is
assumed to be proportional to the number of pixels in which the deposited charge exceeds a defined threshold (`hits'), therefore pixels must be sufficiently small that the multiple-particle probability is negligible even in the core of high-energy
electromagnetic showers.
Due to charge sharing and geometric effects, a single charged particle is likely to generate a group of adjacent hits (a `cluster'). 
To avoid saturation effects originating from overlapping charge clouds and  ensure competitive resolution and linearity, 
 very high transverse granularity sampling is achieved using 
binary-readout CMOS pixels. The small pixel size also has clear benefits
in dense particle environments for pattern recognition algorithms such
as particle flow, see \eg Refs.~\cite{Brient:2001fow,Morgunov:2001cd}.

The first proof-of-concept demonstrations of a DECAL \cite{Ballin:2008db,Ballin:2009yv,Dauncey:2010zz} used custom-designed sensors to show calorimetric behaviour in test beams  as part of R\&D for a future International Linear Collider.
 The first
proof-of-principle of a fully functional  DECAL took place  in the context of the ALICE
experiment forward calorimeter (\focal)
\cite{Focal-loi}, with the design, realisation and measurements of a
multiple-layer prototype proving the viability of this novel approach to calorimetry \cite{deHaas:2017fkf}.
For the \focal detector, the main role of the pixel technology is to provide discrimination between pairs of photons from neutral pion decays and single photons. Although excellent energy resolution is not the main motivation for this application, it is clearly beneficial.

All designs for digital calorimeters use a sandwich structure of silicon and tungsten layers, with Monolithic Active Pixel Sensors (MAPS) providing the necessary high granularity at reasonable cost. The initial DECAL studies developed custom sensors to explore the requirements for such a device, including per-pixel thresholds \cite{Dauncey:2010zz}.
The \mimosa  proof-of-principle prototype \cite{deHaas:2017fkf} required a total sensor area of almost 400~cm$^2$ and therefore used the readily available PHASE2/MIMOSA23 chip from IPHC \cite{Winter:2010zz} with a pixel size of 
\SI[product-units=power]{30 x 30}{\micro\m}. This sensor proved to be well-suited for this first development step but too slow for the future applications envisaged due to the relatively long integration and readout time of \SI[product-units=power]{640}{\micro\s}.

The most recent development step is the \epical prototype presented in this paper. This fully digital pixel calorimeter is constructed primarily to explore the  suitability of the state-of-the-art ALPIDE chip, designed for the ALICE ITS and MFT \cite{ALPIDE}, for digital calorimetry. The project is closely related to an ongoing development of a detector for proton computed tomography \cite{Alme2020} and uses the same technology for the active layers containing the ALPIDE sensors.
As this MAPS sensor is compatible with the interaction rate and data acquisition of the ALICE experiment, the \epical technology is a candidate for the pixel layers of the proposed \focal detector. As the ALPIDE sensors have been qualified for the radiation environment of the ALICE ITS, they are suitable for use in \focal where the expected damage due to radiation load is of similar (or lower) order of magnitude.

It is also a decisive step towards making  digital pixel calorimetry available more broadly for other high-energy physics experiments.
The ALPIDE chip was designed for charged particle tracking and therefore requires the simultaneous measurements of only a small number of clusters from single minimum-ionising particles in a sensor. One of the main questions to be answered using the \epical prototype is  whether the ALPIDE sensors can operate successfully in the high local hit density environment of an electromagnetic shower.

In this paper, the design of this new prototype and its readout, the commissioning and basic performance tests of the device, and the measurement setup and conditions at the DESY test beam will be described. Finally, results of measurements with electron showers will be presented and discussed. More details about the chip properties and much more extensive analyses will be the scope of forthcoming papers.

\section{The \epical prototype}
\label{sec:prototype}

Figure~\ref{fig:design} presents an overview of the 
\epical prototype, a
 silicon-tungsten (Si-W) digital calorimeter that has 24 ultra-high granularity active layers\footnote{A left-handed coordinate system is used throughout, with the $z$-axis pointing downstream and perpendicular to the front face of layer 0 (most upstream), the $y$-axis horizontal, and the origin at the geometric centre of the layer-0 active area.}
instrumented using ALPIDE sensors.
Each layer consists of two ALPIDE chips of 50~$\mu$m thickness plus cabling and 3~mm of tungsten absorber, as shown in  figure~\ref{fig:singleLayer}. The total sensor area per layer is \SI[product-units=power]{30 x 30}{\mm}, out of which \SI[product-units=power]{27.6 x 29.9}{\mm} are sensitive.
This prototype allows the performance of the ALPIDE chip to be characterised in the context of digital calorimeter applications. It also provides input to the final \focal design parameters by testing the corresponding electronics, cabling and readout components.

\subsection{Mechanical construction and cooling}

\begin{figure}[btp]
\centering 
\includegraphics[width=\textwidth,origin=c]{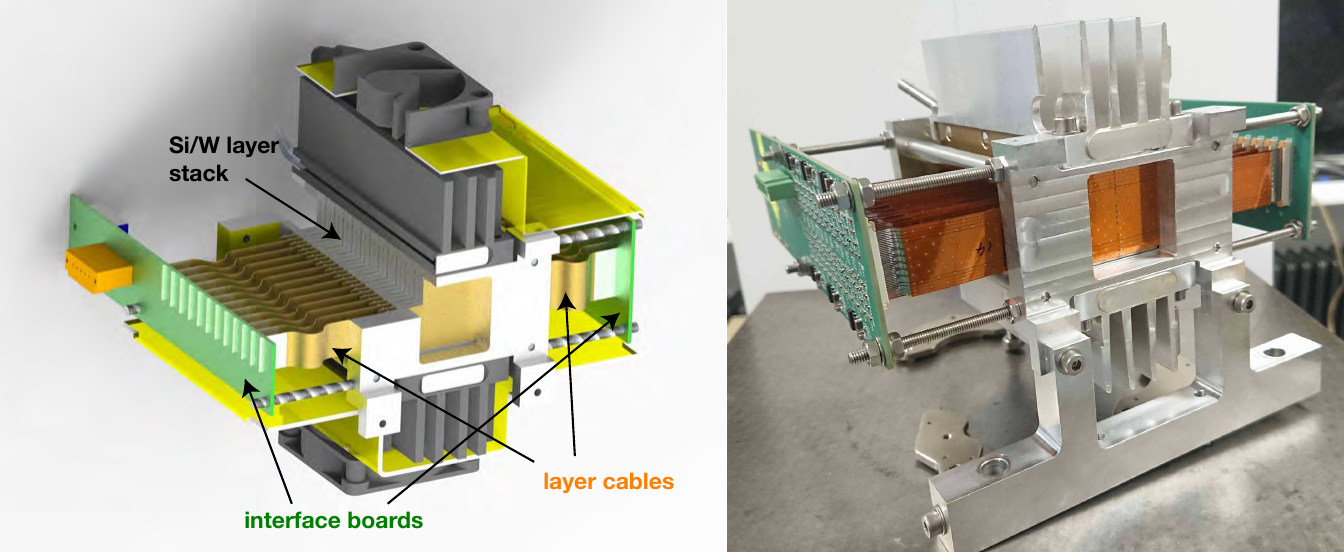}
\caption{\label{fig:mTower} (left) Design drawing and (right) photo of the final construction of the \epical prototype. Layers are oriented such that the most upstream layer (0) has no tungsten absorber in front of the ALPIDE sensors.\protect\label{fig:design}}
\end{figure}
\begin{figure}
    \centering
    \includegraphics[width=\textwidth]{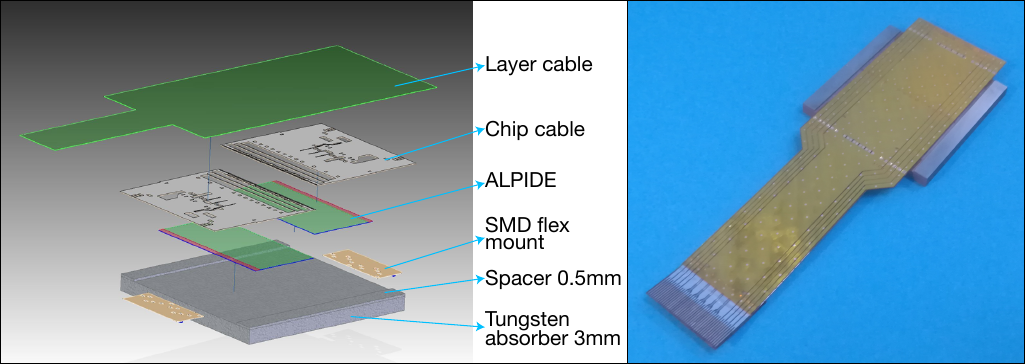}
    \caption{(left) Exploded view of the design of a single \epical prototype layer and (right) photo of an assembled layer. See text for details.}
    \label{fig:singleLayer}
\end{figure}

The \epical layers including tungsten absorber, ALPIDE chips and cabling (see figure~\ref{fig:singleLayer}) were constructed at the Research and Production Enterprise `LTU', Kharkiv, Ukraine where the ultra-thin aluminium-Kapton cabling has been developed and produced.
This prototype demonstrates that mechanical overheads, including those associated with cabling, can be controlled to derive maximum benefit from the small Moli\`{e}re radius of the tungsten absorber.
Two ALPIDE chips are glued  onto an absorber plate with minimal separation between their long edges, resulting in a small gap in $y$ of \SI[product-units=power]{\approx 100}{\micro\m} in the centre of each layer. Each ALPIDE chip is tape-automated bonded to a chip cable, which is itself bonded to a layer cable that  supplies power to the analogue and digital chip circuits, as well as providing a high-speed link for the raw data stream. The ALPIDE power rail SMD (Surface Mount Devices) style decoupling capacitors are
carried by the SMD flex mount, which is also attached to the chip cable.

The 24 layers are stacked on top of one another and housed within an aluminium structure, as shown in figure~\ref{fig:design}, that allows stable temperature to be maintained during operation using a water cooling system.  Air cooling is also possible.

Each complete layer in the \epical assembly is \SI{3.5}{\mm} thick: the tungsten plate has a thickness of \SI{3}{\mm} and a surface of \SI[product-units=power]{40 x 40}{\mm}; two tungsten strips (called `spacers' in figure~\ref{fig:singleLayer}) of width \SI{4}{\mm} and thickness \SI{0.5}{\mm} are placed at two opposite edges of the plate to protect the ALPIDE and cabling in a recessed region when the layers are stacked. The total thickness of ALPIDE and cabling is less than \SI{0.5}{\mm}.

\subsection{ALPIDE chip and readout}
The ALPIDE chips have a size of
\SI[product-units=power]{30 x 15}{\mm}, consisting of a  matrix of $1024 \times 512$ pixels of area \SI[product-units=power]{29.24 x 26.88}{\micro\m}. The readout is hit-driven and each double column of pixels is read out via a priority encoder.
The ALPIDE sensors are supplied with 1.9~V, and the typical current per sensor is 75-100~mA. They are connected via a 40~MHz clock input, a bidirectional control signal and a 1.2 Gbps encoded HSDATA output. The internal signal peaking time is 2~$\mu$s.

\begin{figure}[b]
    \centering
    \includegraphics[width=\textwidth]{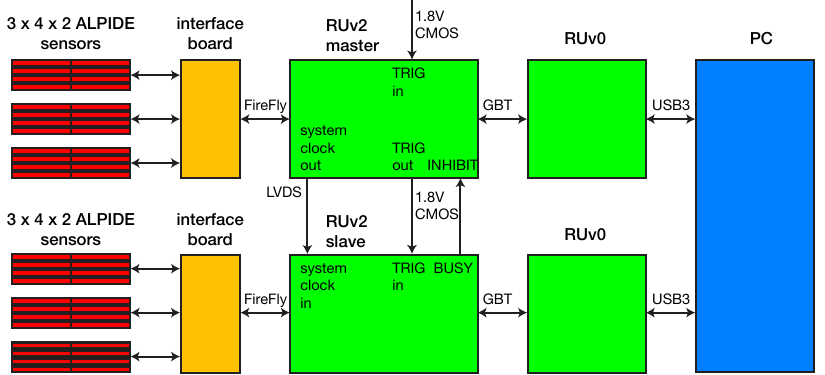}
    \caption{Schematic view of the readout system of the \epical detector.}
    \label{fig:readout}
\end{figure}

A schematic overview of the readout system is shown in figure~\ref{fig:readout}. Twelve layer cables from each side of \epical are connected to an interface board using ZIF connectors (see figure~\ref{fig:design}). The interface board segments the 12 modules in three groups, each with two local regulators (1.9~V ALPIDE analogue and digital power) and a SAMTEC FireFly link for connecting the eight 1.2~Gbps HS-datalines and the shared clock and control lines to the FPGA-based RUv2 (Readout Unit, v2 \cite{8824419,RUv2}). The RUv2 is connected via a GBT \cite{Moreira:2009pem} fibre optic link to a previous design iteration of the readout unit (RUv0), which emulates the Common Readout Unit \cite{Mitra:2016tfb} widely used in the ALICE experiment. Both RUv0 are connected to a PC via a USB3 connection.

\subsection{Data acquisition and trigger}
The data acquisition is based on the Python code developed for the ALICE ITS Upgrade \cite{RUfirmware} and runs on a standard Linux PC.
The RUv2 firmware has been adapted to enable external triggering and readout of more than nine ALPIDE data streams. 
For the readout of the 48 ALPIDEs in the prototype, two sets of RUv2 and RUv0 are needed. 

The prototype readout was triggered using two scintillator tiles placed approximately\ 35~mm upstream of layer~0. A third identical scintillator tile is placed behind the prototype, to allow characterisation using cosmic muons as introduced in section~\ref{sec:cosmics}. The BC420 plastic scintillator tiles are \SI{3}{\mm} thick  and have an area of \SI[product-units=power]{3 x 3}{\cm} to match the sensitive area of the prototype.
Each tile is read out by a low-noise SiPM (Hamamatsu MPPC S13360-1325PE),   surface mounted to a PCB that supports the tile. The SiPM is placed in a dimple in the centre of the scintillation tile, which itself is enclosed by reflective adhesive foil to increase light yield and uniformity of response. The tile plus PCB are covered by  light-tight black tape. These scintillator modules have been developed at MPP Munich for the CLAWS beam-background monitoring system at SuperKEKB/Belle II  \cite{SIPM-MPP}.

The two RUv2 boards run in a master-slave configuration, distributing the system clock for synchronisation to ensure simultaneous data taking. For the test beam, a coincidence signal of the two upstream scintillators was generated using NIM electronics modules and provided as a \SI{1.8}{\V} CMOS signal to the master RUv2 readout board. The master RUv2 sends the trigger signal on to the slave, while a possible BUSY condition of the slave is also taken into account.  
For cosmics measurements, a similar coincidence was set up with the second (front) and third (back) scintillators.

At the beginning of a data-taking run, the DAQ PC requests the RUv0 boards deliver a cycle of $N_{\mathrm{tr}}$ triggers and subsequently checks periodically the number of triggered events  stored on the RUv0. When this number is equal to $N_{\mathrm{tr}}$, triggering is paused and the transmission of the data from RUv0 to the PC is requested.
During this cycle, whenever the RU boards receive a trigger signal, they distribute this to the ALPIDE sensors, which send their local data to the RUv2. These data are stored in FIFO cells on the RU until a cycle has been completed and the data transported to the PC, after which  a new cycle begins. The cycle can also be terminated on exceeding a  configurable time-out $t_{\mathrm{to}}$ within which no triggers are received.
The values of $N_{\mathrm{tr}}$ and $t_{\mathrm{to}}$ can be adjusted according to the data-taking conditions, respecting the size of the FIFO memory on the RU boards.

\section{Experimental setup}
\label{sec:setup}
\subsection{Measurements of cosmic muons}
\label{sec:cosmics}
The \epical setup  was mounted vertically in the laboratory to measure cosmic muons, with two trigger scintillators above and the third below. Cosmic data were taken during a period of about six months in 2020.
Parameters controlling the sensors were optimised during initial data taking and used throughout subsequent measurements.
Due to the very small solid angle of the \epical prototype, the total number of events recorded was approximately 9000.

\subsection{Measurements at the DESY II Test Beam Facility}
\label{sec:testbeam_setup}
\begin{figure}[tbp]
\centering
\includegraphics[width=.75\textwidth]{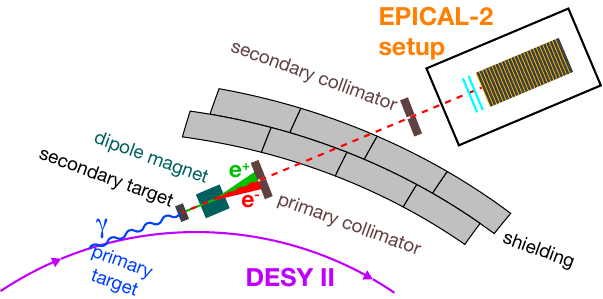}
\caption{\label{fig:tbsetup} The  \epical setup at the DESY II Test Beam Facility  \cite{DESY} (not to scale).}
\end{figure}

In February 2020, the \epical setup was installed in the TB22 beam line at the DESY II Test Beam Facility \cite{DESY}, which provides $e^+$ and $e^-$ beams of nominal momenta $p_0$ between 1.0 and 5.8\gevc (see figure~\ref{fig:tbsetup}). The beam aperture was defined by use of collimator settings. 
Data were collected under varying conditions of nominal particle momentum $p_0$, collimator apertures and incident position of the beam on the prototype, which may also be presented at angles in the range \SI{0}{\degree}--\SI{20}{\degree} relative to the nominal beam direction.
In addition, pedestal data were recorded in the absence of beam to monitor  noise within the detector.

The  linearity of response measured for the prototype depends crucially on knowledge of the momentum scale and its potential variation with $p_0$. The resolution determined also depends on the intrinsic momentum spread of the beam. These characteristics are measured for the adjacent beam line (TB21), with the momentum spread given as $\sigma_p = 158 \pm 6 \mevc$, independent of $p_0$~\cite{DESY}.
Figure~31 in Ref.\ \cite{DESY} presents mean values of measured particle momenta  $p_\mathrm{eff}$ as a function of nominal momentum, showing small deviations from linearity; numerical values corresponding to these data points are summarised in table~\ref{tab:desymomentum} \cite{priv:stanitzki}.
The general trend observed is that the measured momenta are higher than nominal values, with the absolute differences reducing with increasing momentum. The best agreement between nominal and measured momentum is observed for $p_0 = 5\gevc$, which is therefore used as the reference point when interpreting results of this paper.

There is an additional uncertainty related to the highest momentum of 5.6\gevc presented in \cite{DESY}, whereas data herein extend to 5.8\gevc. Although small, this  difference is important as the momenta are close to the phase-space limit of the primary beam, where the production probability is asymmetric and the high energy tail of the particle spectrum is strongly suppressed.
This may lead to a larger uncertainty at 5.8\gevc, which cannot be estimated reliably from the available information.

\begin{table}[bt]
    \caption{Nominal momentum $p_0$ at the TB22 test beam, the corresponding effective  mean momentum $p_\mathrm{eff}$ measured at TB21 \protect\cite{DESY,priv:stanitzki} and the  number of events recorded in the test beam data taking, summed over both electron and positron  beams. Statistical uncertainties are expected to be negligible for $p_\mathrm{eff}$.}
    \centering
    \begin{tabular}{|c|c|c|}
    \hline
    $p_0$~(\gevc)   & $p_\mathrm{eff}$~(\gevc) & $N_{\mathrm{events}}$   \\ \hline
    1.0 & 1.119  & $2.4 \cdot 10^6$ \\ \hline
    2.0 & 2.045  & $2.0 \cdot 10^6$ \\ \hline
    3.0 & 3.026  & $1.8 \cdot 10^6$ \\ \hline
    4.0 & 4.016  & $2.2 \cdot 10^6$ \\ \hline
    5.0 & 4.989  & $3.5 \cdot 10^6$\\ \hline
    5.6 & 5.560  & -- \\ \hline
    5.8 &    --  & $1.5 \cdot 10^6$  \\ \hline
    \end{tabular}
    \label{tab:desymomentum}
\end{table}

The conditions in our measurements differ in terms of beam optics, collimator settings and beam line from those in \cite{DESY}.  As it is not appropriate to correct the \epical  results from nominal to true particle momenta, the differences are therefore used to assign an estimated systematic uncertainty to the beam particle momenta.
The momentum spread according to \cite{DESY} is also applied in simulations to allow  realistic comparisons to data, unless stated otherwise.

Table~\ref{tab:desymomentum} summarises the test-beam data accumulated under stable operating conditions that are used for analysis in this paper. At each energy,
data have been recorded for both $e^{-}$ and $e^{+}$ beams. As no significant difference was observed between the samples, they have been combined at each nominal beam momenta.
 Similarly, a set of nominal settings with respect to the test beam are used, consisting of: apertures of $ 14 \times 14 \, \mathrm{mm}^2$ for the primary, and $ 12 \times 12 \, \mathrm{mm}^2$ for the secondary collimators;
 approximately normal incidence of beam particles on layer~0; 
 and a temperature of the water cooling system of \SI{20}{\degree}~C.

\section{Simulation}
Monte Carlo simulations of the detector response and the shower evolution in \epical are performed using \Allpixtwo \cite{SPANNAGEL2018164}, a generic pixel detector simulation framework based on \geant and ROOT\footnote{\Allpixtwo v1.6.0 with \geant v4.10.07 and ROOT v6.23/01}.
All simulation steps from generation of incident radiation to production of final digitised hits in the readout electronics can be performed within this framework.
Distinct tasks within \Allpixtwo are performed by modules\footnote{Unless noted otherwise, parameters controlling the behaviour of a module remain at their default values.}, with those used to model the \epical prototype described below.
\begin{figure}[t]
    \begin{minipage}[t]{0.55\textwidth}
     \begin{center}
      \includegraphics[width=\textwidth]{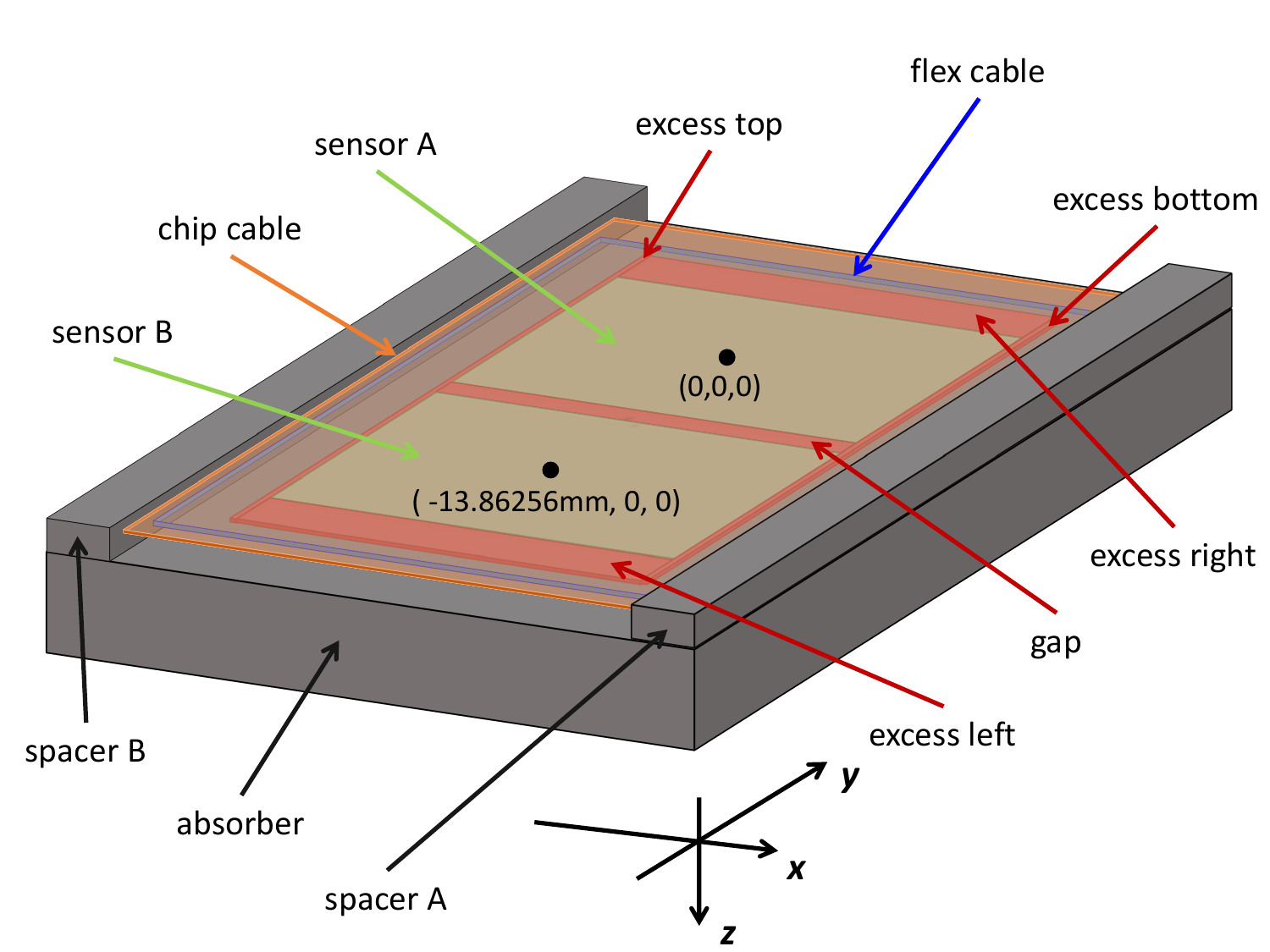}
    \end{center}
  \end{minipage}
  \begin{minipage}[t]{0.45\textwidth}
     \begin{center}
      \includegraphics[width=\textwidth]{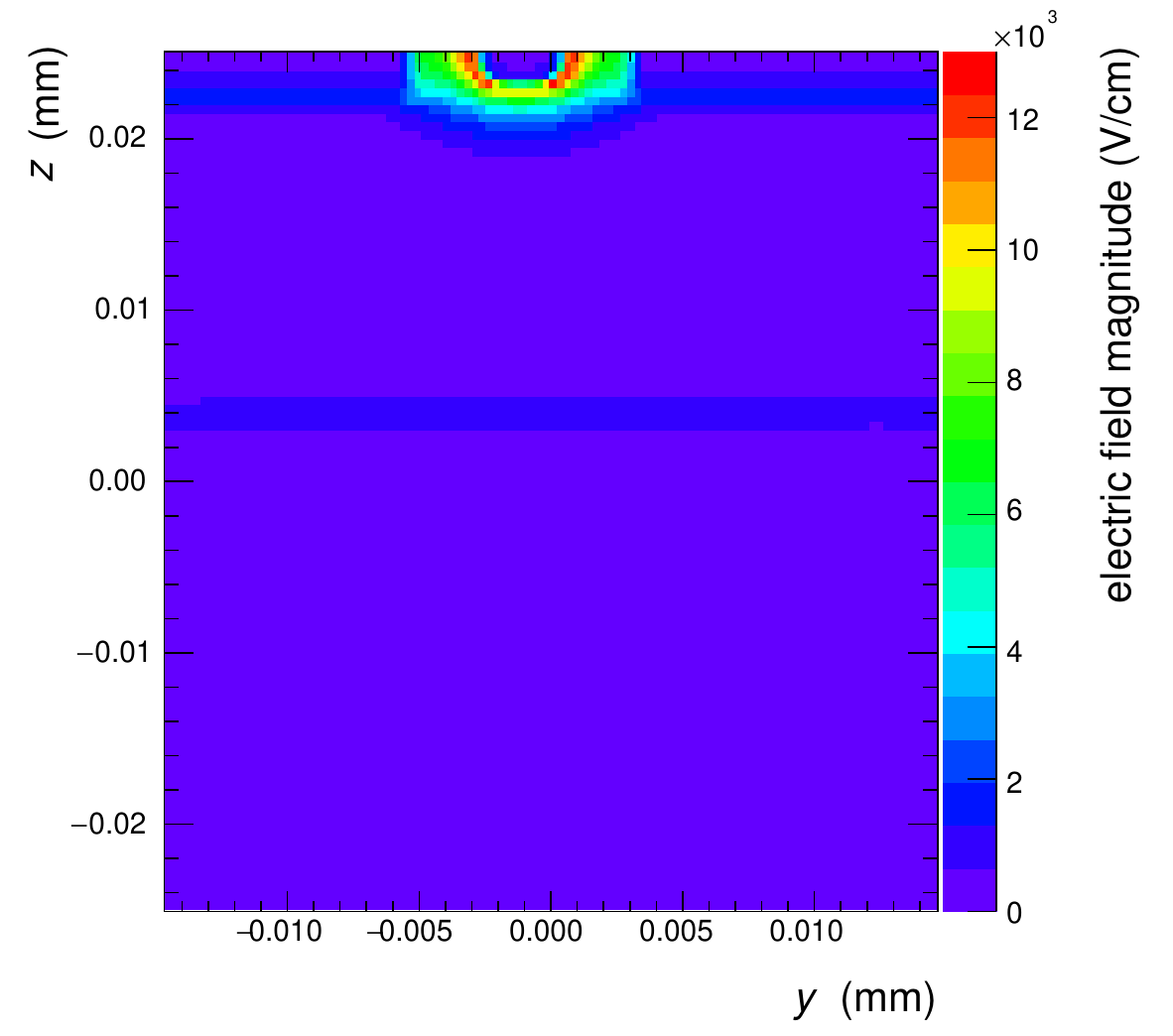}
    \end{center}
  \end{minipage}
  \caption{(left) Implementation of an \epical layer in the \Allpixtwo simulation, with local coordinate system defined (right) Magnitude of the electric field  within a single pixel, in the ($y-z$) plane (sensor local coordinate system).}
  \label{fig:construction}
\end{figure}

The detailed \epical detector geometry introduced in section~\ref{sec:prototype} is defined (module \textit{GeometryBuilderGeant4}) by specifying the type, size, material and position of the detector volumes. The tungsten absorber and spacers are implemented as passive materials as well as the chip and flex cable.
Both active sensors per layer are implemented as 50~$\mu$m thick silicon with a matrix of 1024 $\times$ 512 pixels whereas the pixel size is 29.24~$\mu$m $\times$ 26.88~$\mu$m.
Figure~\ref{fig:construction} shows the implementation of a single instrumented layer in the simulation, with
two sensors separated by a gap of \SI{100}{\micro\metre}.
To account for the guard rings and peripheral circuitry of ALPIDE, in the simulation model the sensors are surrounded by passive silicon, referred to as excess in figure~\ref{fig:construction}.

The electric field is defined for all pixels in each sensor and is added to the detector description (module \textit{ElectricFieldReader}).
 The motion of charge carriers and thus the number of  hits are particularly sensitive to the strength of the electric field, which is therefore  derived from adaptive
TCAD simulations. These model the configuration used during data taking, namely an overall reverse bias voltage on the collection diode of $V_{\mathrm{RB}}=1.4\,$V, which is the sum of the pixel reset voltage ($V_{\mathrm{RST}}=1.4\,$V) and the reverse substrate bias voltage ($V_{\mathrm{BB}}=0\,$V).
Figure~\ref{fig:construction} shows the electric field magnitude for a single pixel as implemented in the simulation.

Particle transport through the whole detector geometry as well as the deposition of charge carriers in the active sensor volumes is performed by the \textit{DepositionGeant4} module, which acts as a wrapper around \geant and uses the selected \geant physics list; FTFP\_BERT\_EMZ was used herein \cite{Bagulya:2017xch}.
As the criteria introduced in section~\ref{sec:event-selection}  select events with particles limited to a  \SI[product-units=power]{16 x 16}{\mm} region centred on layer~0, the electron beam position in simulation is randomly selected from a square surface of the same dimensions and $5\,\text{mm}$  in front of layer~0.
The results obtained with simulations were insensitive to the choice of beam profile, for events that are mostly contained within the prototype.
The energy of the electrons are sampled from a Gaussian distribution with an energy spread of 158\mev, following section~\ref{sec:testbeam_setup}.

All deposited charges are transported through the sensor  by the \textit{GenericPropagation} module, which implements an iterative motion consisting of diffusion (random walk) and drift (user-defined electric field).
Propagating each individual charge carrier provides a microscopically precise simulation on the cost of computing time.
As an alternative to transporting all charge carriers independently, $\Nch=50$ charge carriers are propagated as a set, reducing the computing time substantially without influencing the bulk properties obtained from the simulation.
Above $\Nch=100$, a reduction in cluster size and thus a smaller number of hits is observed.
In addition, charge carriers are propagated in the time frame of a so-called integration time \tint.
The transport of a charge-carrier set terminates either when \tint is exceeded or when the set reaches any surface of the sensor considered.
As \tint is not directly known from the ALPIDE sensor, a value (25.1~ns) is chosen to optimise agreement with 5\gev electron data\footnote{To retain predictive power in the simulation, this was tuned using data only at one beam energy setting.}.
It was observed that longer integration times increase cluster sizes and the number of hits.
Since charge sharing between pixels becomes significant after $\approx 15\,$ns, for values around $\tint = 25.1\,$ns the number of shared charges reaching the collection diode mainly influences the total number as well as the number of nearby pixel hits.

The propagated sets of charges are mapped to their nearest pixel
by the \textit{SimpleTransfer} module.
For each pixel, the mapped sets of charges are combined into a group, ignoring those more than $d_{\mathrm{count}} = \SI[product-units=power]{5}{\micro\m}$ depth from the sensor surface in the pixel implant region.
Finally, the group of charge carriers is processed by the frond-end electronics.

The response of the front-end electronics is simulated by the \textit{DefaultDigitizer} module, where the charges assigned to a pixel are translated into a digitised signal.
Gaussian noise contributions centred on  zero and with standard deviation of  20 electrons are added on an event-by-event basis; this is in the upper limit range and therefore rather too high.

The hit thresholds can be implemented individually for each pixel in the simulation. From threshold scans the mean threshold and the sigma of its sensor-internal variation is known for all sensors. The overall mean value of the thresholds is 82e, and the mean of all sensor-sigmas is 20e. The pixel-by-pixel variation of the threshold appears to be random throughout the sensors. While the shape of these fluctuations is not strictly Gaussian, it does not vary strongly, so for simplicity the actual pixel-by-pixel thresholds in simulation are sampled from a Gaussian distribution of mean 82e and sigma 20e.
The simulated output includes information of every pixel hit in terms of column, row and layer.

\section{Analysis and corrections}

\subsection{Pixel masking}
\label{sec:noise-removal}
In general, two different types of malfunctioning pixels can be identified. 
Noisy pixels are those that produce a signal significantly more often than the majority of pixels under similar conditions.
Dead pixels are those that only intermittently (if ever) show a response to either the passage of a particle or the injection of a test signal directly into the pixel; their inclusion therefore contributes to fluctuations in the apparent signal.
Both noisy and dead pixels have a detrimental impact on the calorimeter performance and are therefore excluded by a pixel mask that can be applied either during data taking or subsequently.
One of the chips in layer~20 was disabled in the data taking due to a readout error and was therefore not taken into account for the pixel masking.

As part of the quality assurance process, all sensors have undergone extensive probe tests including the injection of signals directly into their front-end circuits, where those with excessive power consumption, missing functionality, or an unacceptably large number of malfunctioning pixels were rejected. All accepted sensors are classified from these tests, and this classification is one source of information to identify noisy and dead pixels. This was complemented here with results from pedestal runs and electron beam runs.
Pedestal runs are taken during the test beam period in the absence of a particle beam.  
These data are acquired using an external trigger signal from a function generator while pixel masking is disabled. As no physics signals are expected under these conditions, pixel hits must originate from electronic noise. Pixels showing an excessive rate of hits in the pedestal runs are flagged as noisy.
Electron beam runs also allow poorly behaving pixels to be studied: by averaging over a large data sample, the hit rate in a given pixel is expected to be comparable with those nearby.  Those in which significant deviations are found, and where the beam intensity does not vary rapidly, are identified as either noisy or dead.

In addition to removing noisy and dead pixels as above, a fiducial region is defined to ensure uniformity of response in the analysis, and two areas are excluded by pixel masks as follows. The outermost column and row of each chip show systematically fewer hits than the rest of the pixels; this may be due to the guard ring surrounding the active area limiting the collection of charge from energy deposits on one side of a pixel. There is also one chip with a faulty readout region.

\begin{figure}[btp]
\centering 
\includegraphics[width=0.5\textwidth]{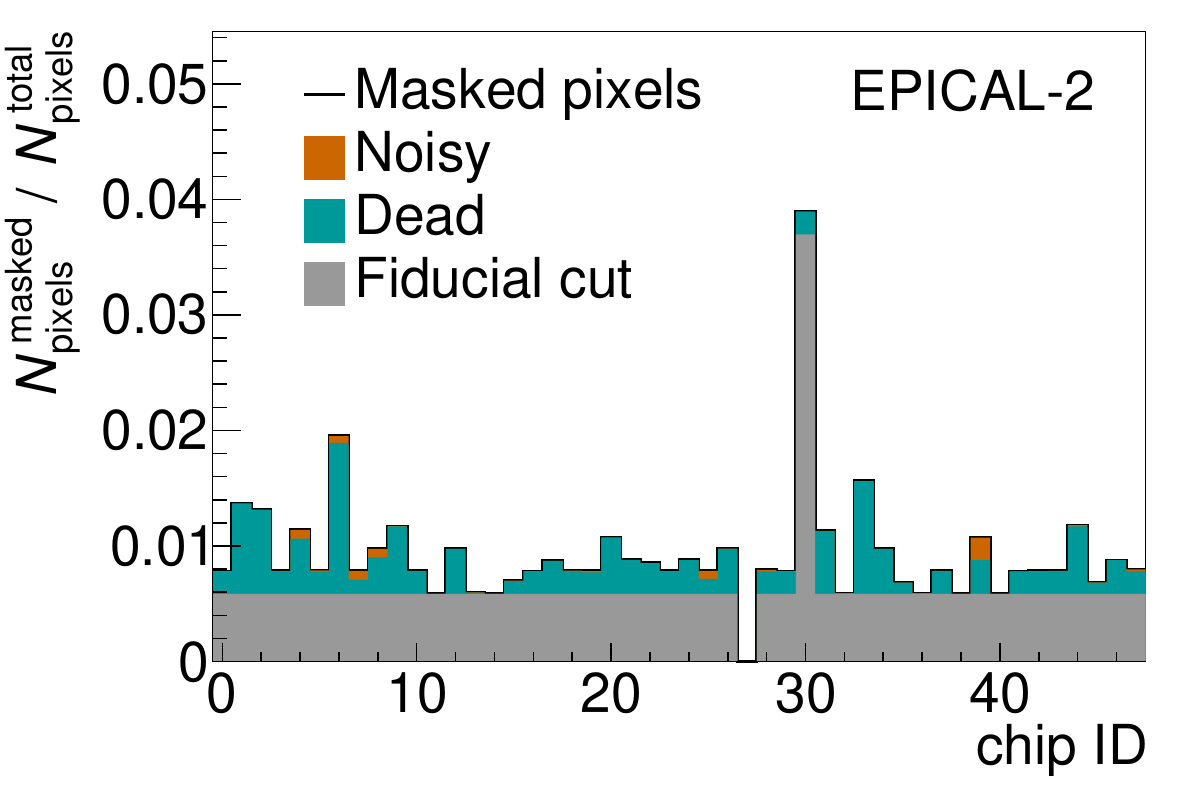}
\caption{\label{fig:pixelmasking:chipdependence} The fraction of all pixels that are masked  for each of the 48 ALPIDE chips (indicated using an arbitrary identifier (ID)).
The fractions of noisy or dead pixels are indicated, as well as those excluded by the  edge-of-sensor fiducial cut described in the text.}
\end{figure}

In total, only 0.95~\% of all pixels are masked, a marginal fraction of 0.015~\% due to excessive noise. 
Dead pixels make up 0.28~\% of all pixels whereas the majority, 0.65~\%, are excluded by the fiducial cut.
Figure ~\ref{fig:pixelmasking:chipdependence} shows the number of masked pixels as a fraction of the total number of pixels as a function of the chip ID.
The overall low level of masked pixels allows for detailed differential shower studies whereas the low chip-by-chip fluctuations demonstrate a uniform performance of the sensors.
The small number of flagged pixels illustrates the very low intrinsic noise levels of the ALPIDE sensors: the relative number of noise hits per pixel of the sensors used is $\approx 4 \cdot 10^{-7}$ without any pixel masking. After applying the mask, this is reduced further to  $\sim 10^{-10}$.

\subsection{Clustering algorithm}
\label{sec:clustering}
The ALPIDE sensors were operated without bias voltage and therefore charge collection at diodes is dominated by diffusion, which increases the charge sharing between neighbouring pixels.\footnote{No bias was applied to the sensors due to the design of the interface boards, therefore charge collection was by diffusion rather than drift. The simpler configuration demonstrates that acceptable cluster size performance is achieved even without bias; this is expected to improve for non-zero bias, further enhancing cluster separation.} Hits are grouped into clusters and both of these quantities are used to characterise the performance of the prototype in terms of its energy resolution and linearity of response.
All pixels in a given sensor use the same threshold setting, which in general differs between sensors.

In our analysis, a cluster is defined as a group of neighbouring pixel hits by one of two algorithms. The default  clustering is based on a geometric pattern search derived from the DBSCAN algorithm~\cite{DBSCAN}. This does not allow clusters to be formed across the insensitive region between two sensors in a given layer.
An alternative clustering scheme is also investigated to reduce computational time for  online use, based on  readout addresses rather than the physical coordinates of pixels. However, this can lead to a small reduction in accuracy relative to the default algorithm, \eg for hits from more than one particle in the same double-pixel  column or where a single particle creates hits in two adjacent region readout units.
\begin{figure}[bt]
\centering 
 \begin{subfigure}[b]{0.475\textwidth}
            \centering
            \includegraphics[width=\textwidth]{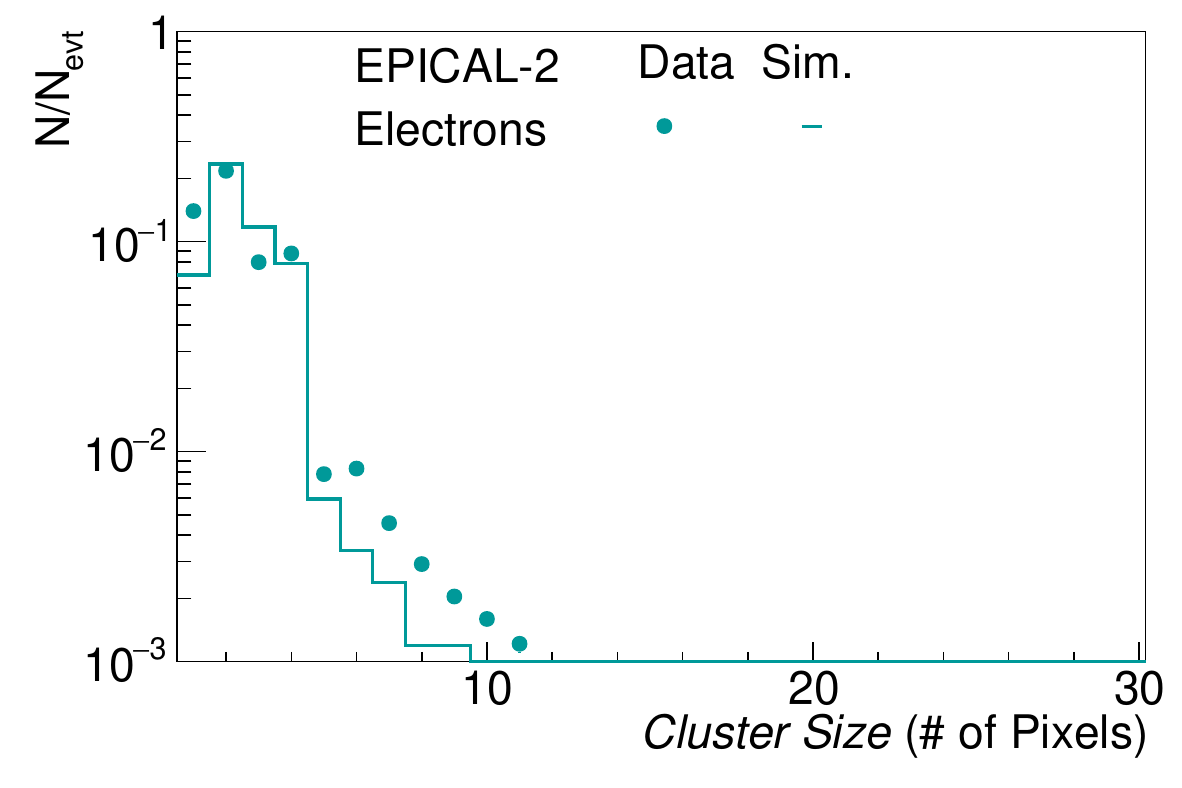}
            \caption{Layer 0}
            \label{fig:clusterSizeLayer0}
        \end{subfigure}
        \hfill
        \begin{subfigure}[b]{0.475\textwidth}  
            \centering 
            \includegraphics[width=\textwidth]{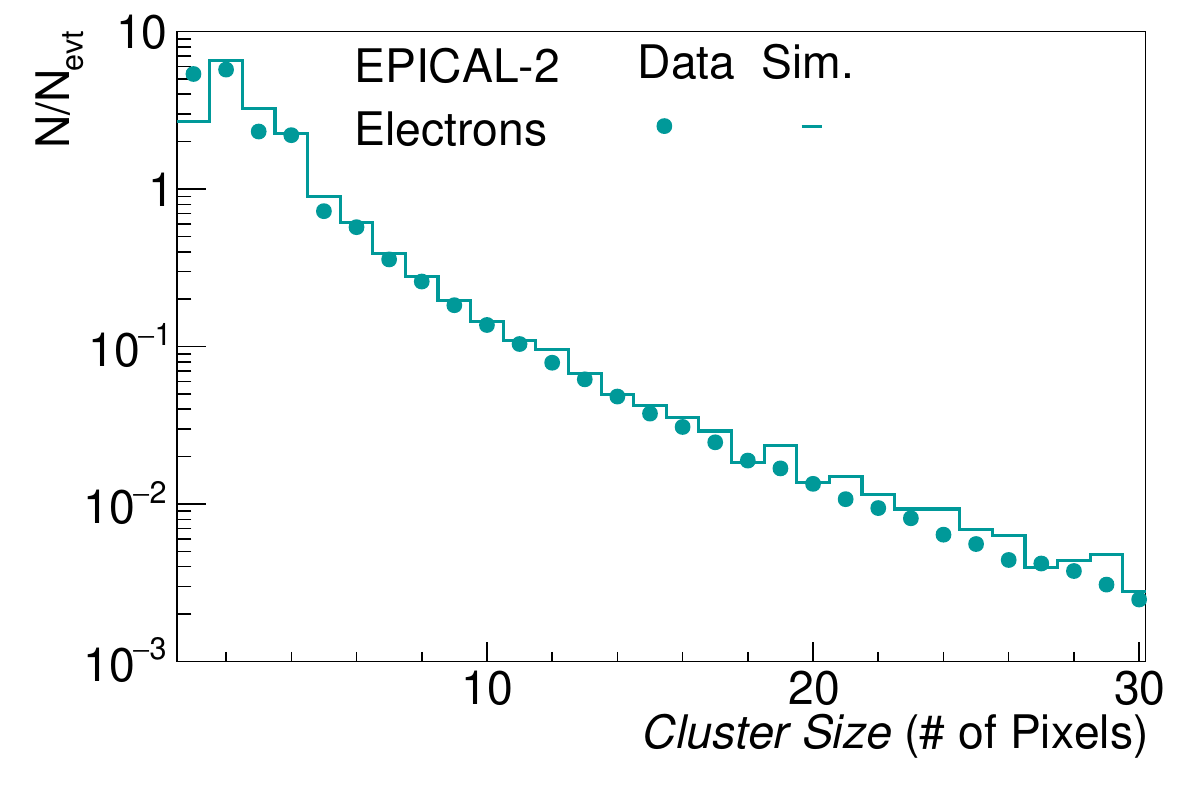}
            \caption{Layer 6}%
            \label{fig:clusterSizeLayer6}
        \end{subfigure}
        \vskip\baselineskip
        \begin{subfigure}[b]{0.475\textwidth}   
            \centering 
            \includegraphics[width=\textwidth]{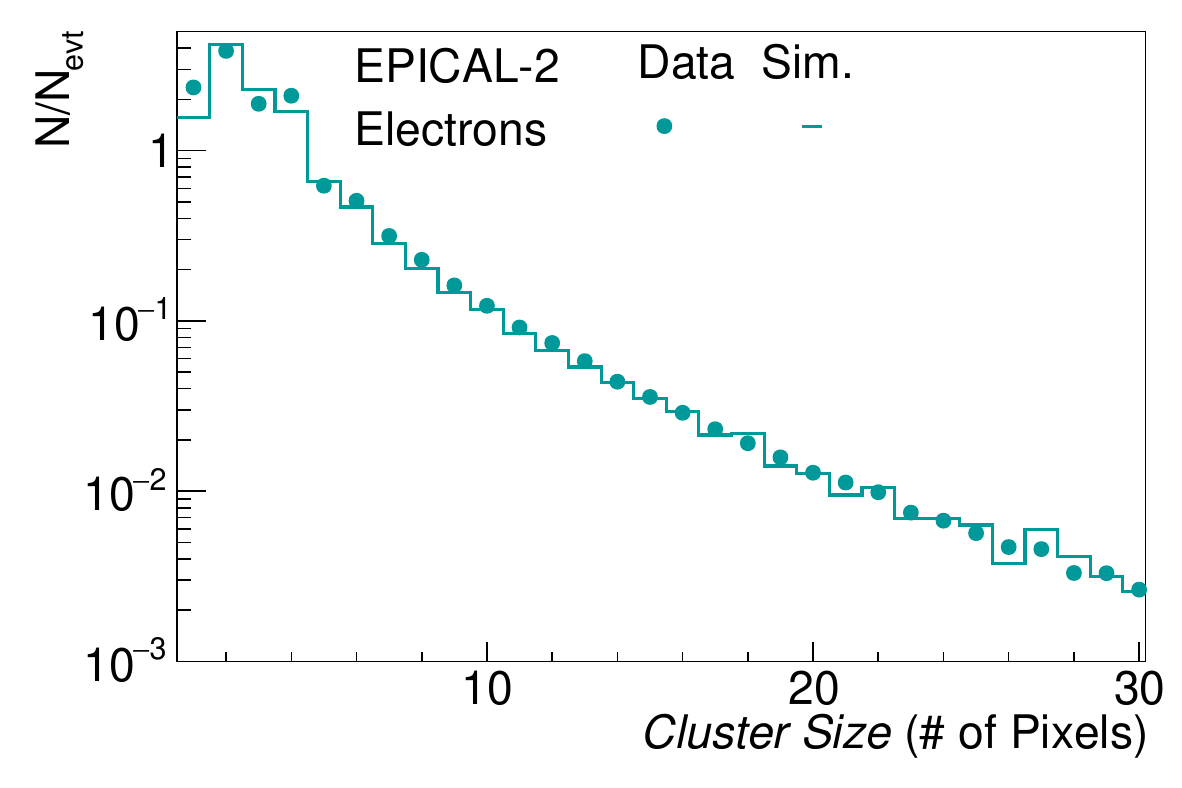}
        \caption{Layer 10}    
            \label{fig:clusterSizeLayer10}
        \end{subfigure}
        \hfill
        \begin{subfigure}[b]{0.475\textwidth}   
            \centering 
            \includegraphics[width=\textwidth]{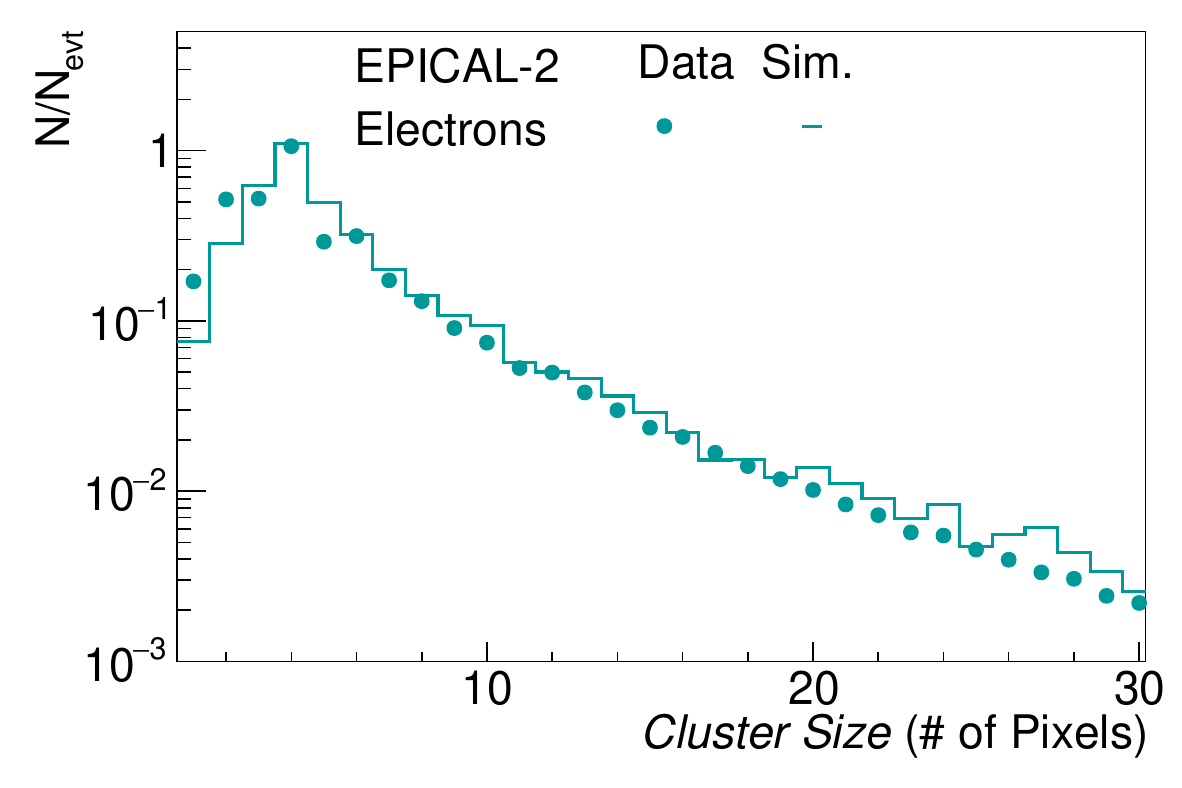}
            \caption{Layer 15}%
            \label{fig:clusterSizeLayer15}
        \end{subfigure}
 \caption{Comparison of distributions of the number of hits per cluster (cluster size) from 5\gev electrons, for data and simulation, see text for details. (Upper left) layer 0, most upstream layer, before any absorber; (upper right) layer 6, approximately shower maximum; (lower left) layer 10; (lower right) layer 15.}\label{fig:clusterSizes}
\end{figure}

\begin{figure}[tb]
\centering 
        \includegraphics[width=0.6\textwidth]{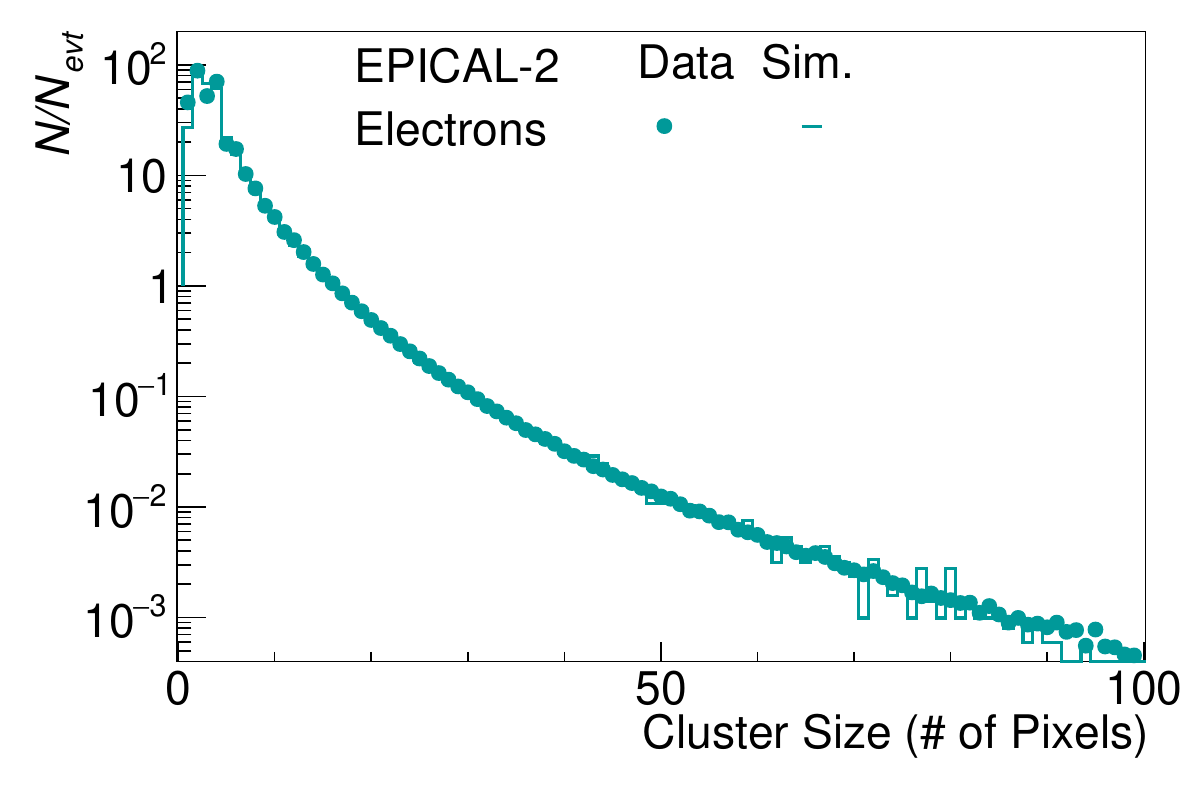}            
        \caption{ Comparison of cluster size from 5\gev electrons, for data and simulation, for all layers.}
      \label{fig:clusterSizesSim}
\end{figure}

\begin{figure}[bt]
\centering
        \includegraphics[width=0.4\textwidth]{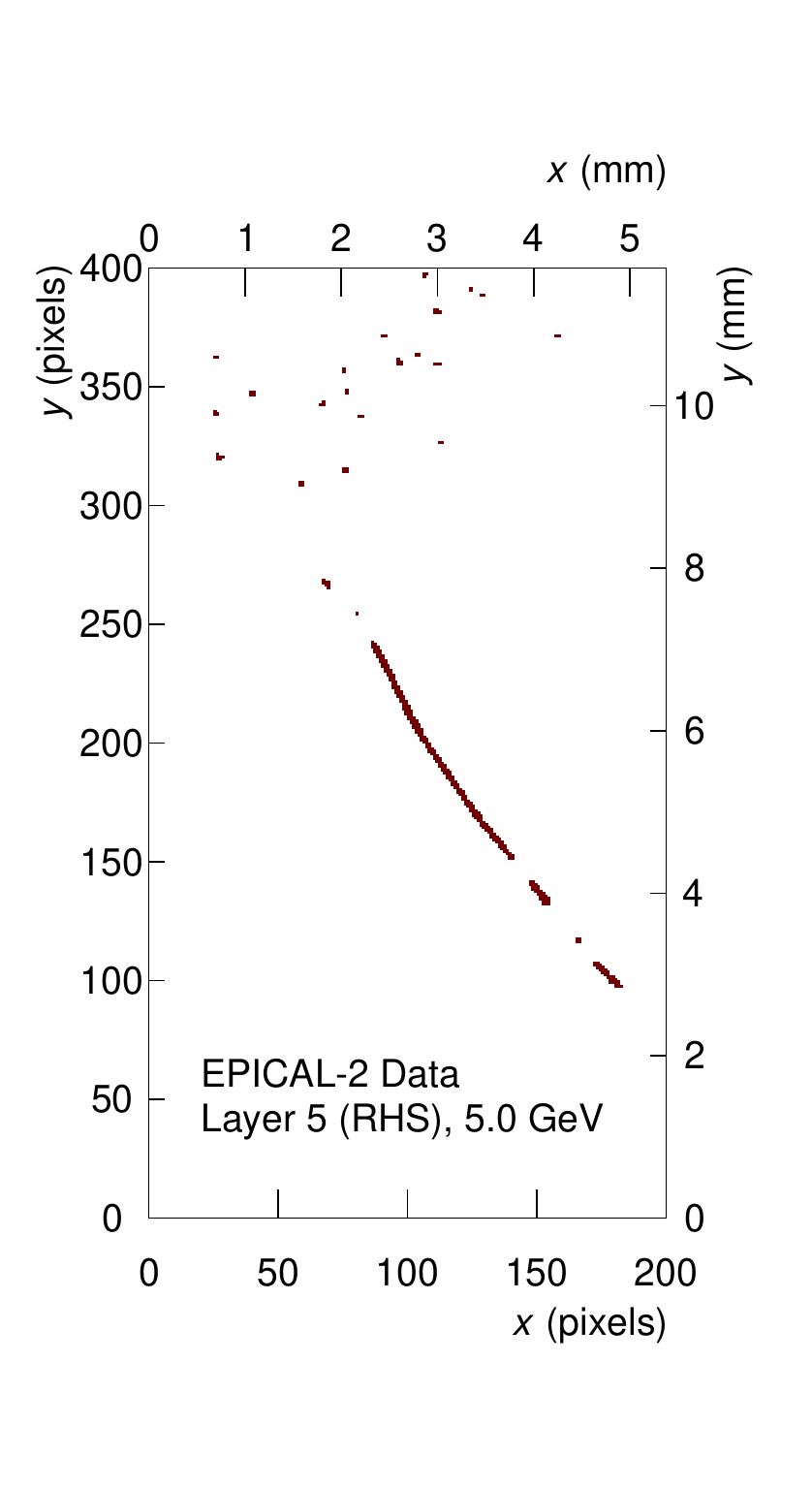}   \includegraphics[width=0.4\textwidth]{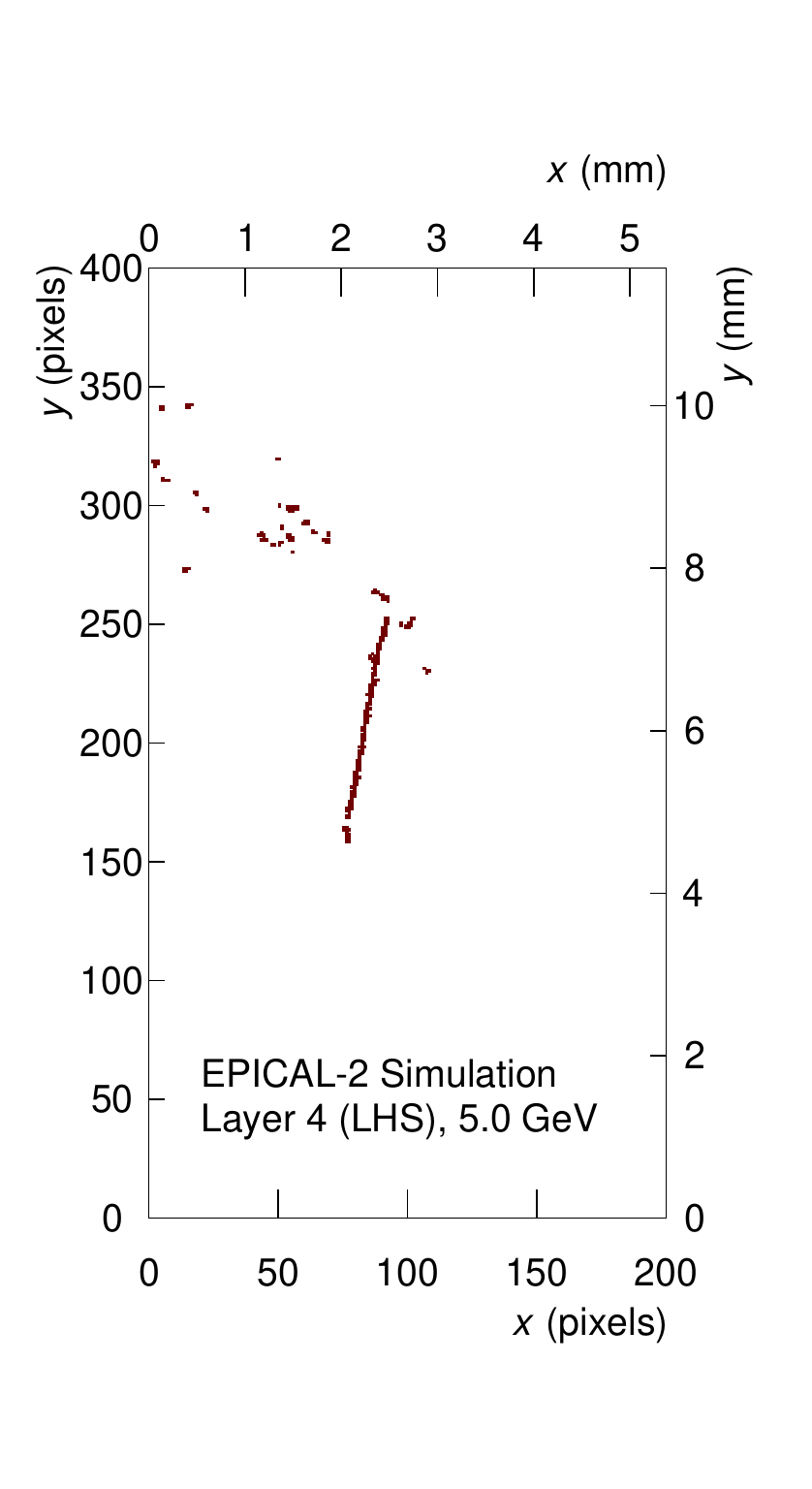}            
        \caption{Single-event displays of hits in a given sensor as a function of $x$ and $y$ position, showing very long, track-like clusters for (left) testbeam data, and (right) simulation, as discussed in the text.}
        \label{fig:longClusters}
\end{figure}

Figure~\ref{fig:clusterSizes} shows the distributions of cluster sizes, \ie the number of hits per cluster, for one chip in each of four layers within the prototype.
Similar qualitative behaviour is found, particularly for the dominant sizes up to four hits, with differences predominantly in the tails of the distributions and the average cluster size.
In layer~0, the cluster size is very small due to the minimal material upstream of the sensor.
Layer~6 corresponds approximately to the longitudinal shower maximum. In this layer,  simulation predicts more single-hit clusters than observed in data, fewer  clusters with between two and six hits, and similar behaviour for very large cluster sizes.
Layers~10 and 15 follow a similar trend, with the latter showing the most probable cluster size of four pixels in both simulation and data, illustrating the quality of the modelling.
The differences observed are dominated by single-hit clusters. Although $\approx 0.95$~\% of pixels have been masked, it is not viable to mask all the pixels that reported hits in pedestal runs. 
The difference in the number of clusters with larger sizes in layer 0, and in general with between two and six hits,  is attributed to the difference in the pixel threshold values of the ALPIDE chips and the MC simulation model. These differences are small enough to be negligible for the calorimetric response.

Figure~\ref{fig:clusterSizesSim} shows the cluster size distribution using 5\gev electron beam data from all layers. 
The probability for either two- or four-hit clusters is in general higher than that for one or three hits, and this is assumed to be due to geometric reasons.
The simulation also models the data well in the region of very large clusters, extending to $\sim10^2$.
The overall features of the distributions are very well described over several orders of magnitude in probability. It is in particular noteworthy that the simulation does predict the existence of very large clusters, which are therefore examined in more detail.
Figure~\ref{fig:longClusters} shows examples of the spatial distribution of hits in one layer having a notably large cluster. An elongated, track-like structure is seen in both data and simulation.  This type of topology dominates in large clusters studied.
In the simulation, it is found that these track-like structures are generated by electrons travelling parallel to the sensor surface.
The observation that the simulation reproduces these rare phenomena so well gives additional confidence in the predictive power of the model.

\subsection{Alignment}
\label{sec:alignment}
\begin{figure}[tb]
\centering
\begin{minipage}[t]{.5\textwidth}
  \centering
  \includegraphics[width=\linewidth]{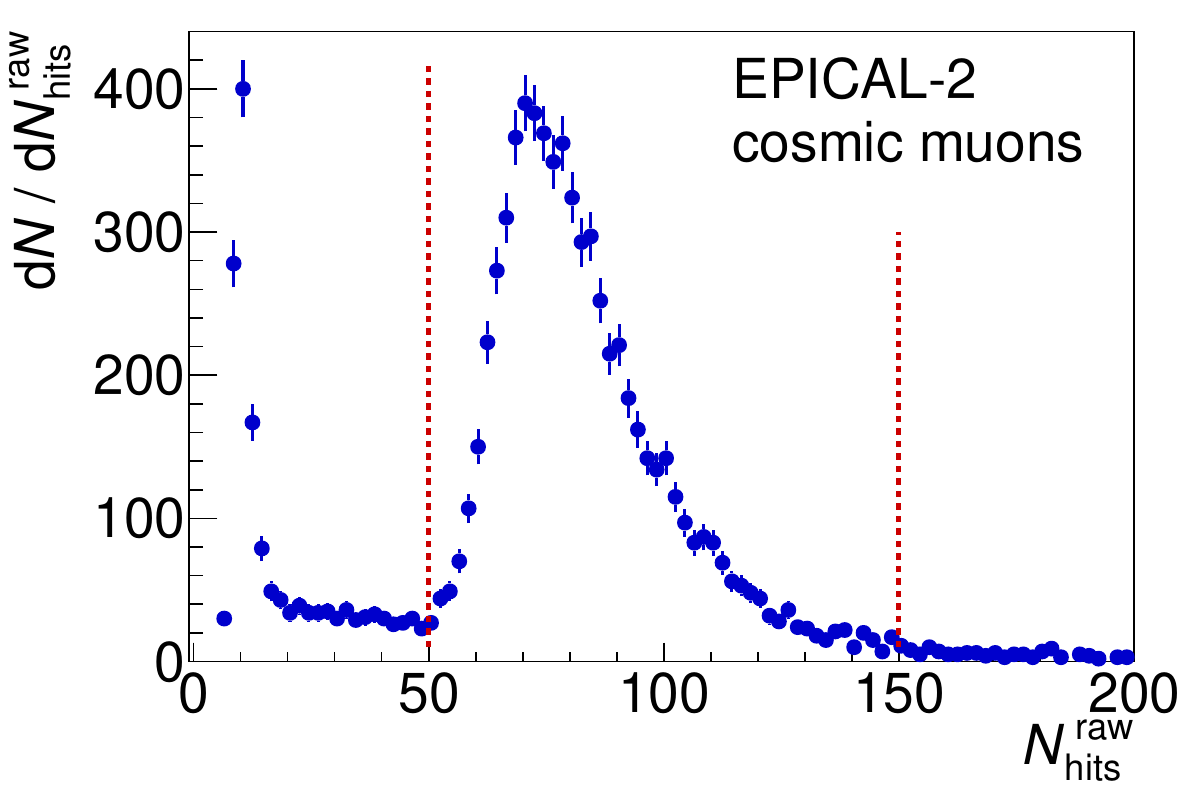}
\end{minipage}%
\begin{minipage}[t]{.5\textwidth}
  \centering
  \includegraphics[width=\linewidth]{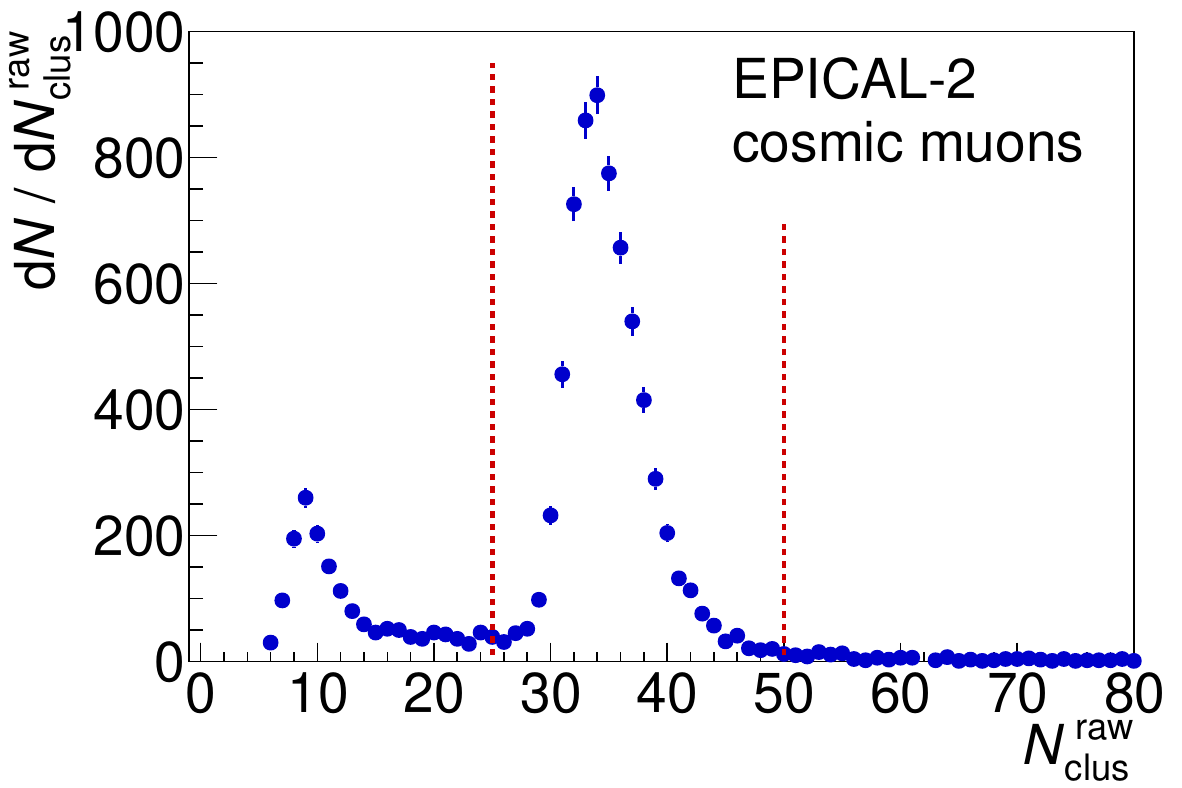}
\end{minipage}
  \caption{\label{fig:cosmic_dist} Distribution of the raw number of (left) hits and (right) clusters, for cosmic muon events. The dashed lines indicate the intervals use for selection of good events.}
\end{figure}

To make efficient use of the small pixel size, the position of the sensors must be known to at least a comparable precision. As this is below that which can be obtained from survey during construction, a data-driven alignment is necessary.
The alignment is performed using tracks of cosmic muons as measured in the laboratory. As the event selection used during alignment requires a straight line track to be reconstructed, additional hits from noisy pixels are not problematic, therefore pixel masking is not used in this analysis
to ensure the required high efficiency, given the low rate of muons.
Figure~\ref{fig:cosmic_dist} shows the distributions of the total number of hits $\Nhits^{\textrm{raw}}$ and clusters $\Nclus^{\textrm{raw}}$ for the data sample. In both distributions, a narrow peak at a value of $\approx 10$ is observed, arising from noisy pixels that create mainly single-hit clusters. The main peak of muons is seen around values of $\Nhits^{\textrm{raw}} \approx 75 $ and $\Nclus^{\textrm{raw}} \approx 34$, with an asymmetric shape including a tail towards large values, reminiscent of a Landau distribution. Cosmic events are selected to have a total number of pixel hits $50 \leq \Nhits^{\textrm{raw}} \leq 150$ and a total number of clusters $25 \leq \Nclus^{\textrm{raw}} \leq 50$, as indicated by the dashed lines in the figure.

The track finding algorithm used is as follows:
\begin{enumerate}

\item In every event, all pairs of hits in different layers are combined to \textit{tracklets}. The extrapolated position of these in layer 0 $(x_i^{\mathrm{zmin}}, y_i^{\mathrm{zmin}})$ and layer 23  $(x_i^{\mathrm{zmax}}, y_i^{\mathrm{zmax}})$ is required to be within the sensor area  (\SI[product-units=power]{30 x 30}{\mm}). The number of retained tracklets is denoted $N_{\mathrm{tl}}$

\item A region of interest (ROI) is defined on both the first and the last layer, consisting of a circle of radius $R_{\mathrm{ROI}} = 0.5 \, \mathrm{mm}$ that contains the maximum number of crossing points from tracklets. All tracklets that do not cross either of the two ROIs are discarded.

\item For every remaining tracklet $i$, a distance measure to another tracklet $j$ is calculated as
\begin{displaymath}
l(i,j) \equiv \sqrt{\left(x_i^{\mathrm{zmin}}-x_j^{\mathrm{zmin}} \right)^2 + \left(y_i^{\mathrm{zmin}}-y_j^{\mathrm{zmin}} \right)^2 + \left(x_i^{\mathrm{zmax}}-x_j^{\mathrm{zmax}} \right)^2 + \left(y_i^{\mathrm{zmax}}-y_j^{\mathrm{zmax}} \right)^2},
\end{displaymath}
and from that the number of neighbours is obtained as 
\begin{displaymath}
N_{\mathrm{n}} (i) = \sum_{j}^{l(i,j) < l^{\mathrm{max}}}, 
\end{displaymath}
with $l^{\mathrm{max}} = 1/\sqrt{2} \, \mathrm{mm}$. The tracklet with the largest value of $N_{\mathrm{n}} (i)$ is chosen as the \textit{track seed} $i_{\mathrm{seed}}$, with the \textit{seed score} 
\begin{displaymath}
S_{\mathrm{seed}} = \frac{N_{\mathrm{n}} \left( i_{\mathrm{seed}} \right)}{N_{\mathrm{tl}}}.
\end{displaymath}

\item All hits in every layer at a relative distance $R \leq R_{\mathrm{h}} = 1 \, \mathrm{mm}$ from the track seed are assigned to the seed. Data from a layer are not used if it includes hits from both sensors, or it has more than one cluster, or it contains a cluster with more than four hits.
The numbers of layers that contribute with valid hits is called $N^{\mathrm{layer}}_{\mathrm{seed}}$.

\item Events are retained for further analysis if $S_{\mathrm{seed}} > 0.98$, $N_{\mathrm{tl}} > 200$ and  $N^{\mathrm{layer}}_{\mathrm{seed}} \geq 18$.

\end{enumerate}

Approximately 4900 events with a muon track candidate are selected from the data sample and used in the alignment procedure introduced below. Track parameters for every event are obtained from a $\chi^2$ minimisation of a straight line to the hit distributions of the associated clusters. 

All sensors in the prototype have six degrees of freedom in their placement: three spatial coordinates and three rotation angles. In the alignment discussed here, it is assumed that the orientation of the layer planes and the longitudinal position in the stack, \ie two angles and the $z$-position are very strongly constrained by the thickness of the tungsten absorber plates and the spacers. Therefore, only two spatial coordinates $(x, y)$ and the rotation angle about the $z$-axis, \ie all possible displacements in the layer planes, are considered as free parameters. Furthermore, with only the relative position information from the detector itself and no fixed absolute reference positions, the alignment procedure is not sensitive to a shear deformation of all layers, \ie a displacement in $x$ or $y$ linearly dependent on the $z$ coordinate. However, as a possible global shear is well-constrained from the mechanical structure, and small shear angles have no appreciable impact on the the detector performance, these are neglected.
The position and angle of two sensors are therefore fixed, one in the most upstream and one in the most downstream layer, to remove this ambiguity.
As the two sensors in each layer effectively form two separate detector stacks, and the majority of the cosmic muon tracks do not cross from one half to the other, the alignment procedure is performed in several phases.

\begin{enumerate}
    \item All sensors in either of the two halves are aligned relative to each other using only tracks that are entirely within in the same half of the prototype.
    \item The alignment of one half of the prototype is fixed, and the other half is aligned relative to the first using tracks crossing from one half to the other.
    \item The alignment of all sensors is updated simultaneously using all tracks.
\end{enumerate}

\begin{figure}[htbp]
\centering
\includegraphics[width=0.9\textwidth]{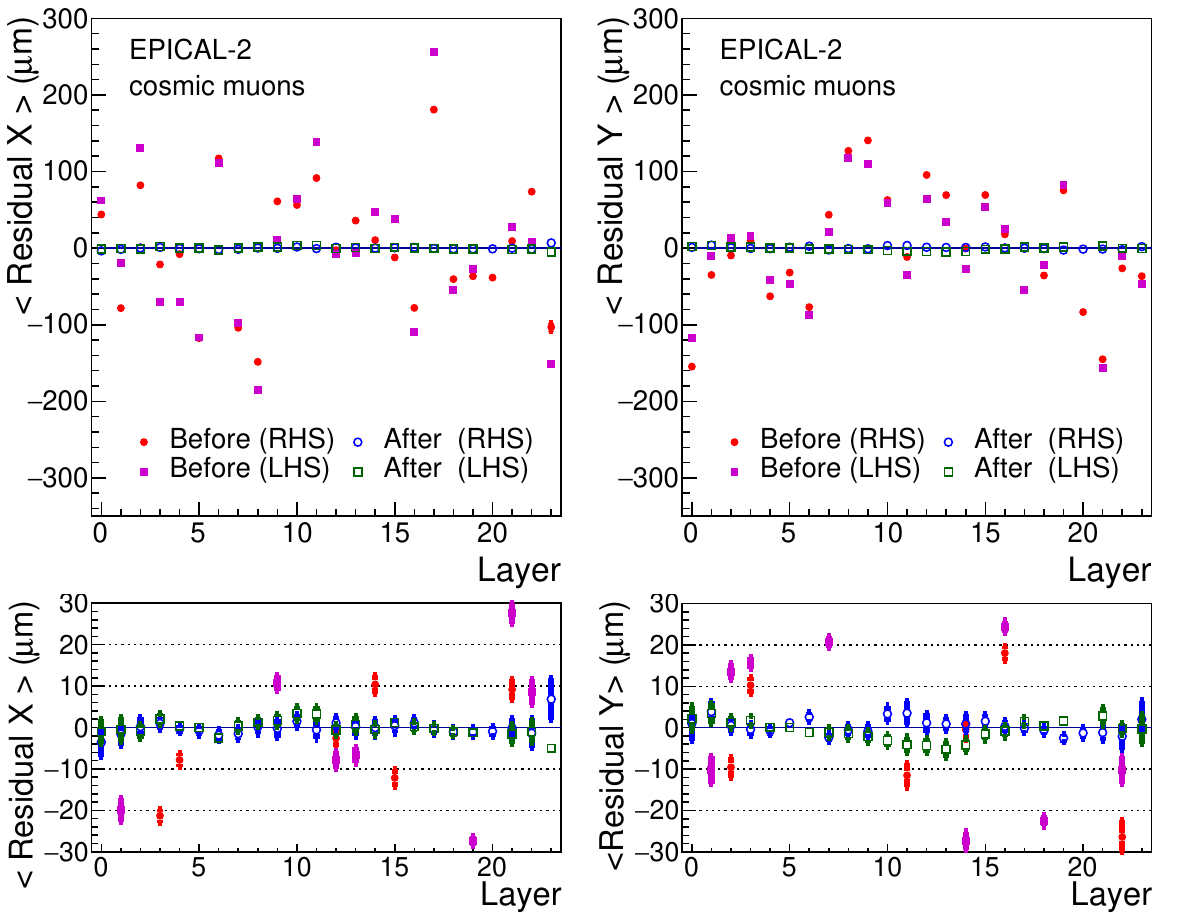}
\caption{\label{fig:residuals} Comparison of the mean values of the track residuals in each layer,  before and after alignment, for the left- and right-hand sides of the prototype. The residuals are shown separately for (left) $x$ and  (right) $y$. The lower panels present the same data within a reduced range to illustrate the alignment achieved.}
\end{figure}

In all phases, the $\chi^2$ of the track per event and from that a global $\chi^2$ averaged over all events used, is calculated. The alignment parameters of all free sensors are then varied iteratively in all three phases until the parameters are stable.

Figure~\ref{fig:residuals} shows the mean values of the residuals of the cosmic muon tracks before and after alignment, for both the $x$- and $y$-directions. Before alignment, the residuals have values up to a few hundred \SI[product-units=power]{}{\micro\m} , which is consistent with the expected assembly tolerances. After alignment, the residuals are of the order of \SI[product-units=power]{5}{\micro\m} or smaller. This level of alignment is sufficient for the proper usage of the \epical prototype.

For the test beam measurements, the relative positioning and orientation of the detector cannot be achieved with a similar accuracy. For the positioning this is uncritical, as the position of incidence of individual beam particles can be reconstructed very well.
However, there is the possibility that the actual global angle of incidence of the particle beam differs  from the nominal one. As this could potentially distort the lateral shower distributions discussed in section~\ref{sec:lateral}, in particular for layers at larger depths, this must be taken into account.

For small angles, a rotation is effectively a shear deformation as discussed above, and corresponds to a lateral displacement of the showers in more downstream layers relative to the first. The displacement can be determined very accurately using an event-by-event analysis of the residuals between the centroids of the hit distributions in the different layers. All layers yield a displacement consistent with global angles in the $x$-$z$ and $y$-$z$ planes. A rotation angle of less than $0.5^{\circ}$ is found in both cases and a linear fit of the displacements determines the correction for each layer. 

\subsection{Event selection}
\label{sec:event-selection}
As the trigger scintillators have approximately the same transverse area as the \epical and are separated in $z$, some beam particles may produce a trigger signal and also generate a shower that is not fully contained in the transverse plane of the prototype. Similarly, for some data recorded the beam intensity was sufficiently high that more than one particle  might enter the calorimeter within the same readout cycle.  As the trigger configuration did not provide rejection of such events, selection criteria are applied offline to remove both these and also partially contained events.
Two algorithms have been developed to select events, one based on the application of a series of sequential criteria (`cut-based') and the other derived from a jet-finding algorithm \cite{Cacciari:2008gp}. 

In the cut-based algorithm, the transverse positions of clusters in the first three layers of the detector are compared, and it is assumed the beam enters perpendicular to layer 0.
A candidate consists of clusters in at least two layers that have transverse positions within 300~$\mu$m of each other.
Candidates are removed if they are within 600~$\mu$m of another candidate that includes a cluster in layer~0.  Where a pair of candidates without clusters in layer~0 are within 600~$\mu$m of each other, they are combined into a single candidate.
The event is accepted if there is only one remaining candidate, otherwise it is rejected.
Further fiducial requirements are imposed on the transverse position of the candidate: it must be at least 8~mm away from all detector edges, and all clusters in layer~1 must be within a 3.6~mm radius of the candidate.

Although differing in their efficiencies for retaining single-electron events and for rejecting both partially contained and multiple-electron events, both selection algorithms give comparable results for the \epical prototype performance, illustrating the stability of results.
For the analysis reported in this paper, the jet-finding algorithm described below is used  as it provides a more stringent rejection of multiple-electron events, as discussed in section~\ref{sec:SelectionResults}.

\subsubsection{Jet-Finding Algorithm}
A grid of transverse cluster positions is defined with  cells of size 
\SI[product-units=power]{0.5 x 0.5}{\mm} in the central part of the detector ($\lvert x \rvert <  \SI[product-units=power]{12}{\mm}$ and $\lvert y \rvert <  \SI[product-units=power]{12}{\mm}$), and larger cells of size \SI[product-units=power]{1 x 1}{\mm} in the outer part ($\lvert x \rvert \geq \SI[product-units=power]{12}{\mm}$ and $\lvert y \rvert \geq \SI[product-units=power]{12}{\mm}$). 
This is to optimise the algorithm for performance in the central, fiducial region of the prototype given the need to reduce leakage  in the transverse plane.
Cells $c_i$ containing clusters from at least three different layers of the detector are associated with the parameters of a pseudojet $i$, according to:
\begin{align}
    \Nclus^c & \to k_{t}^{i} \textrm{ (transverse momentum)}\;\;, \\
    x^{c} &\to y^{i} \textrm{ (rapidity)}\;\;, \\
    y^{c} & \to \phi^{i}\textrm{ (azimuth)}\;\;, 
\end{align}
where  $\Nclus^c$, $x^c$, $y^c$ represent the number of clusters, and the transverse positions of a given cell $c$, respectively.
The anti-k$_{\textrm{t}}$ jet-clustering algorithm \cite{Cacciari:2008gp} is applied to the resulting pseudo-jets, using $R=\SI[product-units=power]{0.5}{\mm}$. Any jets constructed using a single pseudo-jet are rejected as they are very unlikely to be the result of an electron entering the prototype.
The event is only accepted for further study if a single jet, which is expected for an electromagnetic shower from a single incident electron, is obtained.
To ensure containment of the shower, the jet is required to be within a central \SI[product-units=power]{16 x 16}{\mm} fiducial region in the transverse plane.

Further requirements are imposed to ensure a well-understood data sample.
Firstly, if any cluster in the two most upstream layers contributing to the jet has a cluster more than 1~mm away from the jet position, the event is rejected.
Secondly, if any two clusters in the two layers that were both involved in the identified jet are more than 0.5~mm away from each other, the event is rejected. These two criteria reduce contamination from events with two overlapping electromagnetic showers, which would produce a single jet using the anti-k$_{\textrm{t}}$ algorithm. Lastly, any events in which no clusters from layer~0 contributed to the identified  jet are rejected.

\begin{figure}[tbp]
\centering
\begin{minipage}[t]{.45\textwidth}
  \centering
  \includegraphics[width=\linewidth]{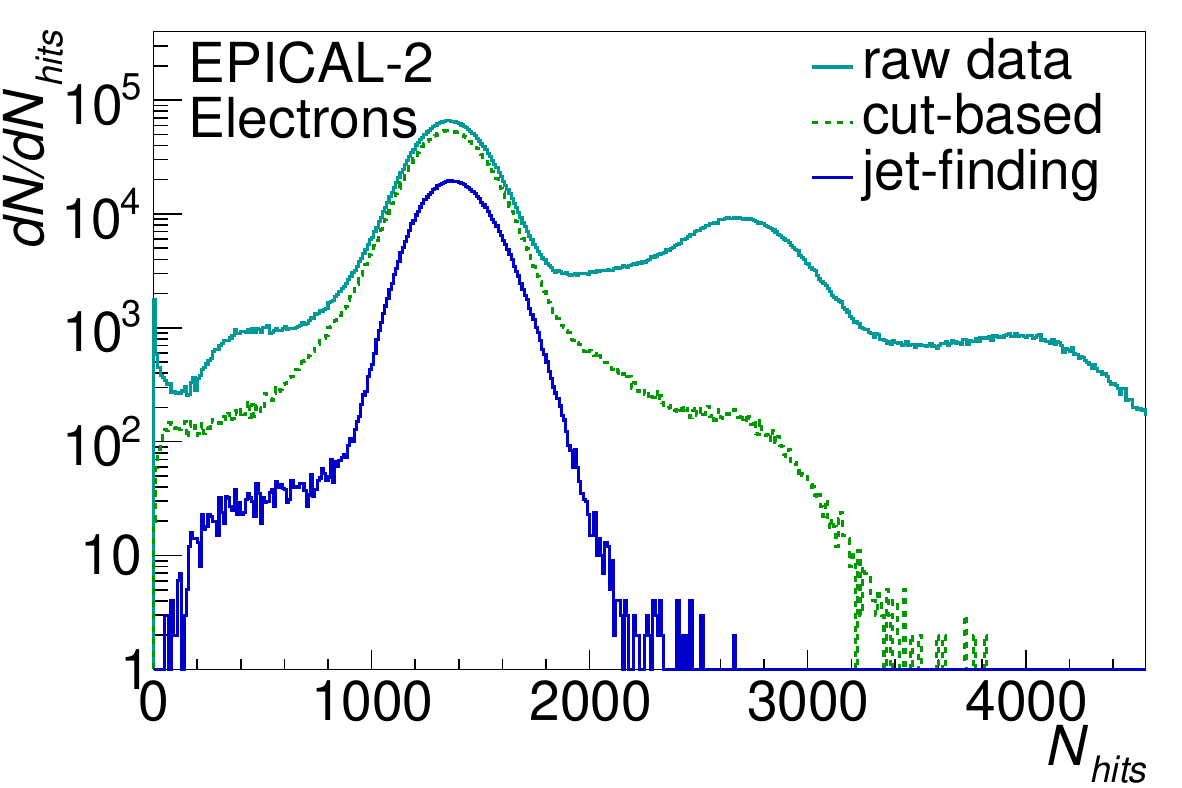}
\end{minipage}%
\hspace*{5mm}
\begin{minipage}[t]{.45\textwidth}
  \centering
  \includegraphics[width=\linewidth]{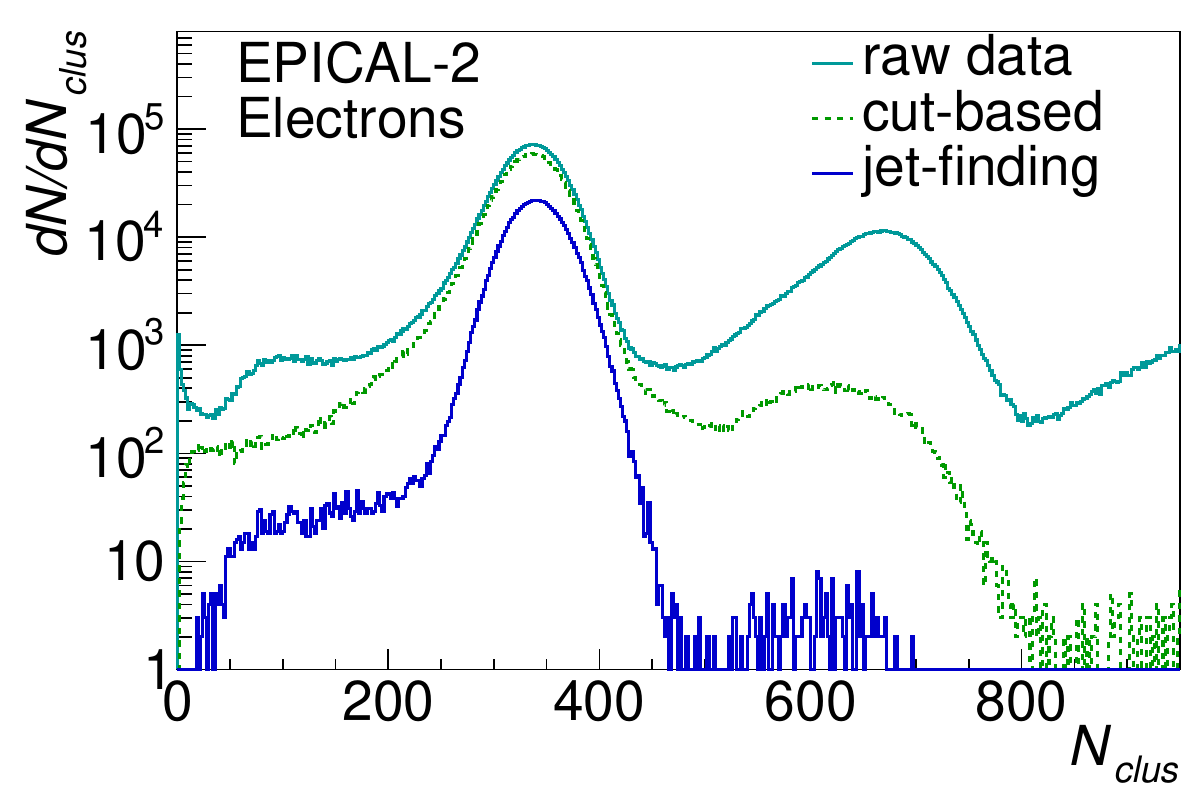}
\end{minipage}
  \caption{Distributions of (left) \Nhits  and (right) \Nclus for 5\gev electrons, before and after event selections are applied, see text for details.}
  \label{fig:nhitsclusterseventselectionlogy}
\end{figure}

\subsubsection{Event Selection Results}
\label{sec:SelectionResults}
Figure~\ref{fig:nhitsclusterseventselectionlogy} shows the distributions of \Nhits  and \Nclus  remaining after each of the selection algorithms for 5\gev electrons.
 Before selections are imposed,  significant peaks are observed at values of $\Nhits  \approx 2700$ and $\approx 4100$, corresponding to events with two and three particles, respectively. Equivalent features are clearly present in the distribution of \Nclus. With the jet-finding algorithm applied, no events remain in the $\Nhits  \approx 4100$  region, and in the $\Nhits \approx 2900$ region the peak is reduced to less than  0.1~\% of the primary peak. The low-intensity shoulder to the left of each distribution has been investigated extensively and is produced by events having small single electromagnetic showers. These showers are consistent with being caused by infrequent low-energy electrons in the beam, possibly produced from interactions with the beam collimators.

Although the distribution of events produced by the jet-finding algorithm is very clean, it has a relatively low efficiency, retaining 21~\% of events. As such, the cut-based algorithm is used as a complementary selection that retains 62~\% of events in the primary, single-electron peak albeit with greater contamination from two-electron candidates.

A small deficit is found in the number of events in a region close to the geometric centre of the detector, which is apparent in the transverse position of hits in the upstream layers of the \epical prototype. This is confirmed as being due to a small inefficiency of the trigger scintillators, and is more pronounced in events where only one scintillator was used. There is no correlation with the distribution of \Nhits inside and outside the affected  region and  no non-uniformity  introduced by this effect in the event-by-event hit distributions. The results presented here are therefore insensitive to this inefficiency.

Beyond this trigger inefficiency, there are potential non-uniformities due to the use of two sensors per layer, which introduces a small insensitive region. To study this quantitatively, the variation of the response to the electron beam with incident particle position has been studied.
To minimise the influence of the finite size of the detector, the response has been limited to a  cylinder of $3 \, \mathrm{mm}$ radius,
\begin{equation}
    \Nhits^{<3 \, \mathrm{mm}} \left( y_{\mathrm{beam}} \right) = \sum_{r < 3 \, \mathrm{mm}} n_{\mathrm{hits}}(r) ,
\end{equation}
where $n_{\mathrm{hits}}(r)$ is the local number of hits at a distance $r$ from the beam particle impact position ($x_{\mathrm{beam}}$, $y_{\mathrm{beam}}$) as determined from the first layer. The relative response is then defined as 
\begin{equation}
    R_{\mathrm{hits}}^{<3 \, \mathrm{mm}} \left( y_{\mathrm{beam}} \right) = \frac{ \left\langle \, \Nhits^{<3 \, \mathrm{mm}} \left( y_{\mathrm{beam}} \right)\, \right\rangle}{\left\langle\left\langle \, \Nhits^{<3 \, \mathrm{mm}} \, \right\rangle\right\rangle}.
\end{equation}
Here, the single angle brackets in the numerator indicate an average over all events, while the double angle brackets in the denominator include a second average over all different horizontal beam positions $y_{\mathrm{beam}}$.

\begin{figure}[tb]
\centering
  \includegraphics[width=0.6\linewidth]{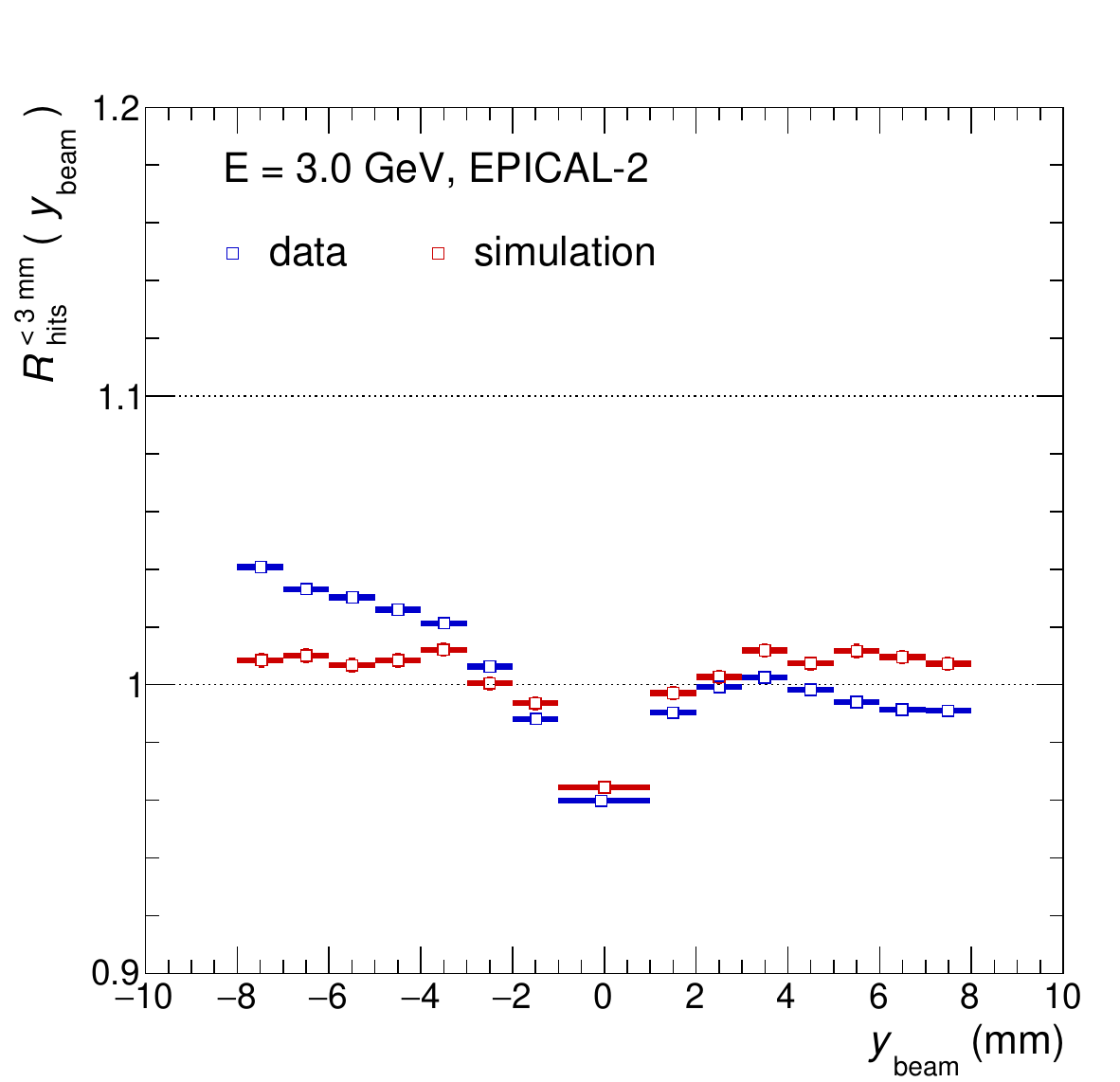}
  \caption{\label{fig:NhitsVSbeamPos} Relative response as a function of horizontal beam position for 3 GeV.}
\end{figure}

Figure~\ref{fig:NhitsVSbeamPos} displays the relative response $R_{\mathrm{hits}}^{<3 \, \mathrm{mm}}$  as a function of the horizontal beam position $y_{\text{beam}}$ for both data and simulation for a 3\gev electron beam. A lower response is found in both simulation and data around $y_{\text{beam}} \approx 0 \; \text{mm}$, which is clearly a result of the inactive gap between the two ALPIDE chips.
In contrast to simulation, the data also show a response variation of about five percent as a function of the horizontal position, arising from horizontal beam energy dispersion at DESY, which is not implemented in the simulation.

\subsection{Calibration}
\label{sec:calibration}
The sensitivity of a sensor is influenced by operating conditions, including  
supply voltage, temperature and discriminator settings in the analogue front-end circuit.
Since these conditions can in principle differ for all sensors, the relative sensitivities between sensors must be calibrated
to ensure unbiased performance. 
In this analysis, cosmic-muon tracks are used for calibration, 
under the assumption that they deposit the same mean energy in each sensor through which they pass as energy loss by the muons is negligible.

A cosmic-muon track is identified following the procedure introduced in section~\ref{sec:alignment}, with the single modification that one layer is excluded from the definition of the track, and the intersection of the track with this layer is determined.
The number of hits and clusters within a circle of radius 3~mm around this intersection are considered as the response of a cosmic muon.
Although multiple interactions of a cosmic muon lead to several clusters in $\approx 7$~\% of all events, most cosmic-muon tracks produce a single cluster with fewer than five pixel hits in a given chip.

\begin{figure}[tbp]
\centering
\begin{minipage}[t]{.5\textwidth}
  \centering
  \includegraphics[width=\linewidth]{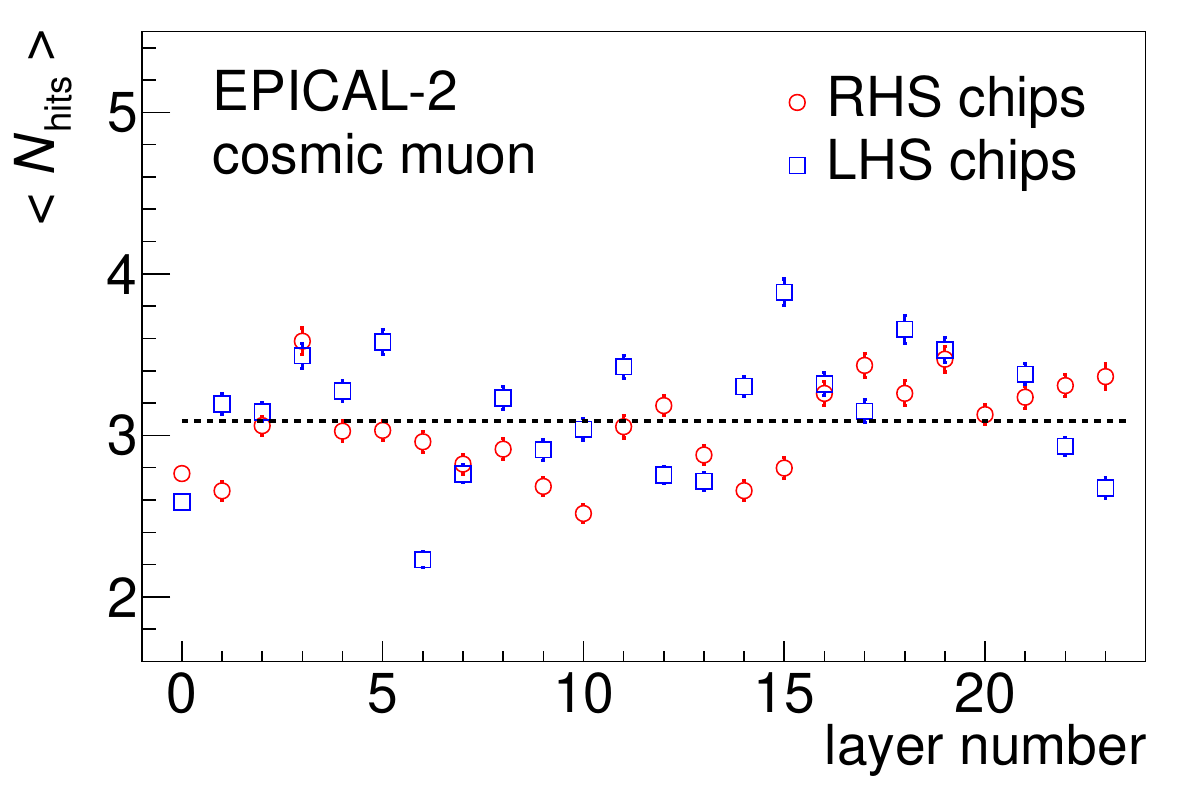}
\end{minipage}%
\begin{minipage}[t]{.5\textwidth}
  \centering
  \includegraphics[width=\linewidth]{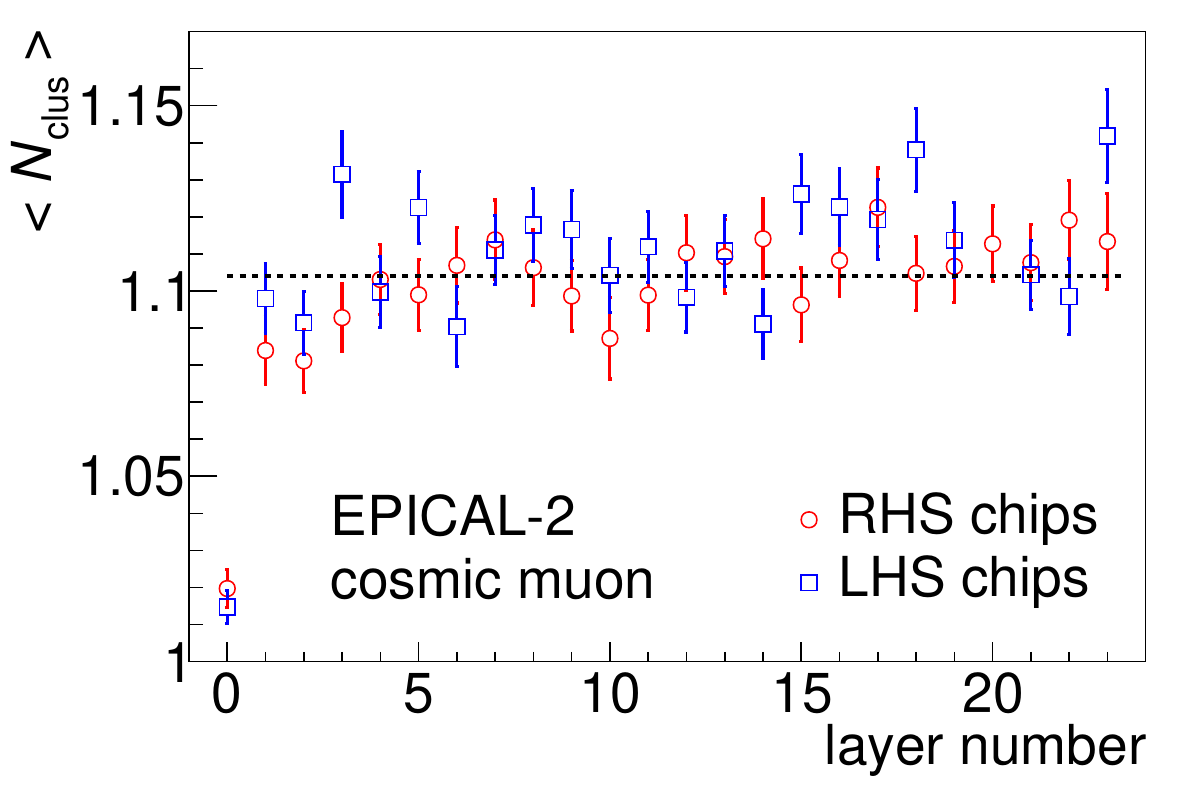}
\end{minipage}
  \caption{Dependence as a function of layer number of the mean number of (left) hits and (right) clusters, originating from cosmic-muon tracks. In each layer, the results are shown separately for chips on the  left-hand and right-hand sides of the prototype. }
  \label{fig:calibration_avergae_nhits_nclus}
\end{figure}

The mean hit and cluster response is calculated separately for each chip, from the corresponding distribution of \Nhits or \Nclus  of the selected cosmic-muon tracks.
Figure~\ref{fig:calibration_avergae_nhits_nclus} shows these mean values separately for both chips in all layers 0--23, together with the average value over all chips.
The larger deviation between chips for $\langle \Nhits\rangle$ than that for $\left< \Nclus \right>$ implies that the cluster measurement is not  affected significantly by operating conditions.

A correction factor is determined per chip for hit (cluster) measurements as the ratio of $\langle\Nhits\rangle$ ($\langle\Nclus\rangle$) to the average value over all chips.

\section{Performance results}
\label{sec:performance}
\subsection{Energy response and linearity}
\label{subsec:response}
The incident electron energy is measured by counting the number of charged particles in each electromagnetic shower.  Two alternative estimators for the number of such particles are considered, namely \Nhits and \Nclus.  
Both quantities provide a measure of the number of charged particles crossing the sensor layers and implicitly of the total track length of charged shower particles.
The number of hits produced by a single charged particle will be subject to statistical fluctuations that depend on charge sharing between adjacent pixels and hence their geometry.
With an appropriate algorithm to define clusters, these fluctuations should be reduced.
As shown in section~\ref{sec:clustering}, the average cluster size will  depend on the individual values of the sensor thresholds and therefore  varies sensor by sensor. The highly uniform response among pixels \cite{ALPIDE} enables each ALPIDE sensor to have a single threshold, which has operational benefits. For energy deposits in the sensors that are well separated, clusters are expected to be the preferred response variable 
as they are less susceptible to these fluctuations and therefore allow better energy resolution to be achieved.

In a high particle density environment, as expected in at least the core of an electromagnetic shower, clusters will start to overlap and, depending on the exact clustering algorithm, may give a biased estimate of the response. The most straightforward effect expected is that of saturation: when more than one charged particle generate nearby clusters, they may be identified as a single large cluster rather than several separate clusters. In this situation, clusters have a disadvantage compared to hits. Although ultimately the number of hits will also be affected by such overlap effects, it is expected that the bias from saturation on the number of hits is significantly smaller than on the number of clusters.

\begin{figure}[tb]
\centering
\begin{minipage}[t]{.45\textwidth}
  \centering
  \includegraphics[width=\linewidth]{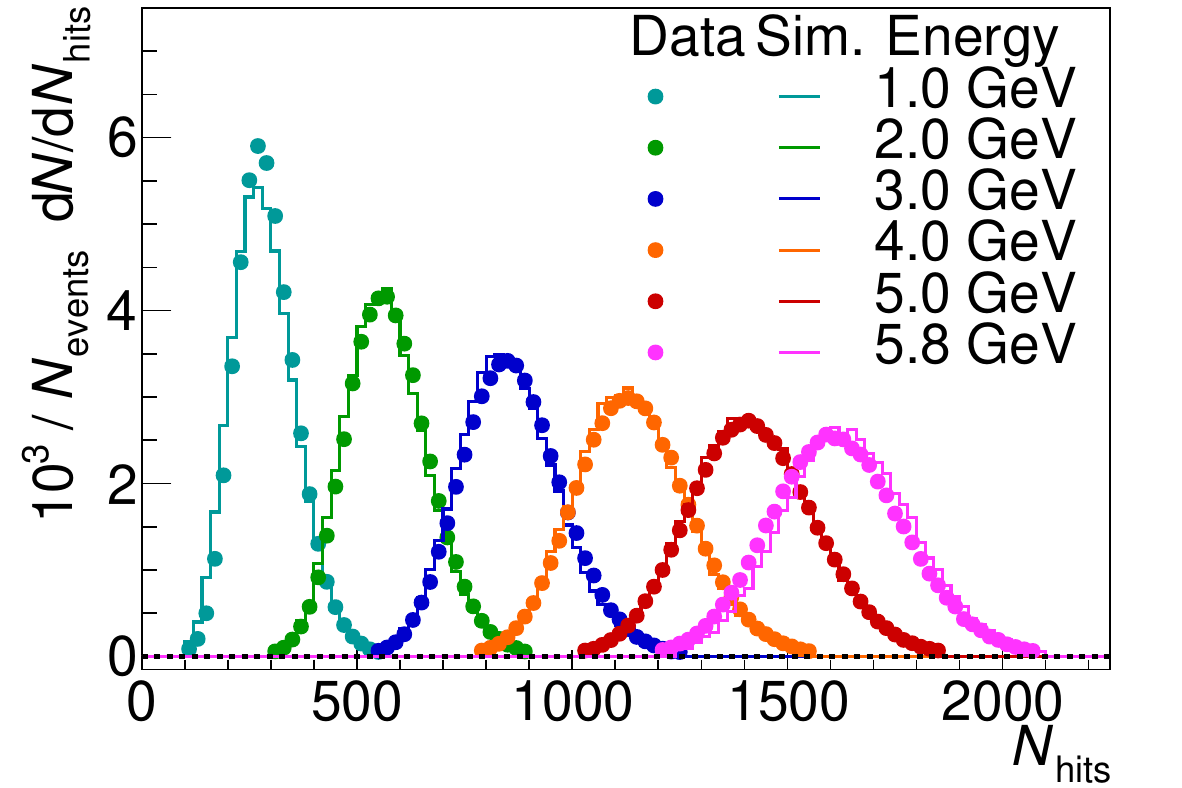}
\end{minipage}%
\begin{minipage}[t]{.45\textwidth}
  \centering
  \includegraphics[width=\linewidth]{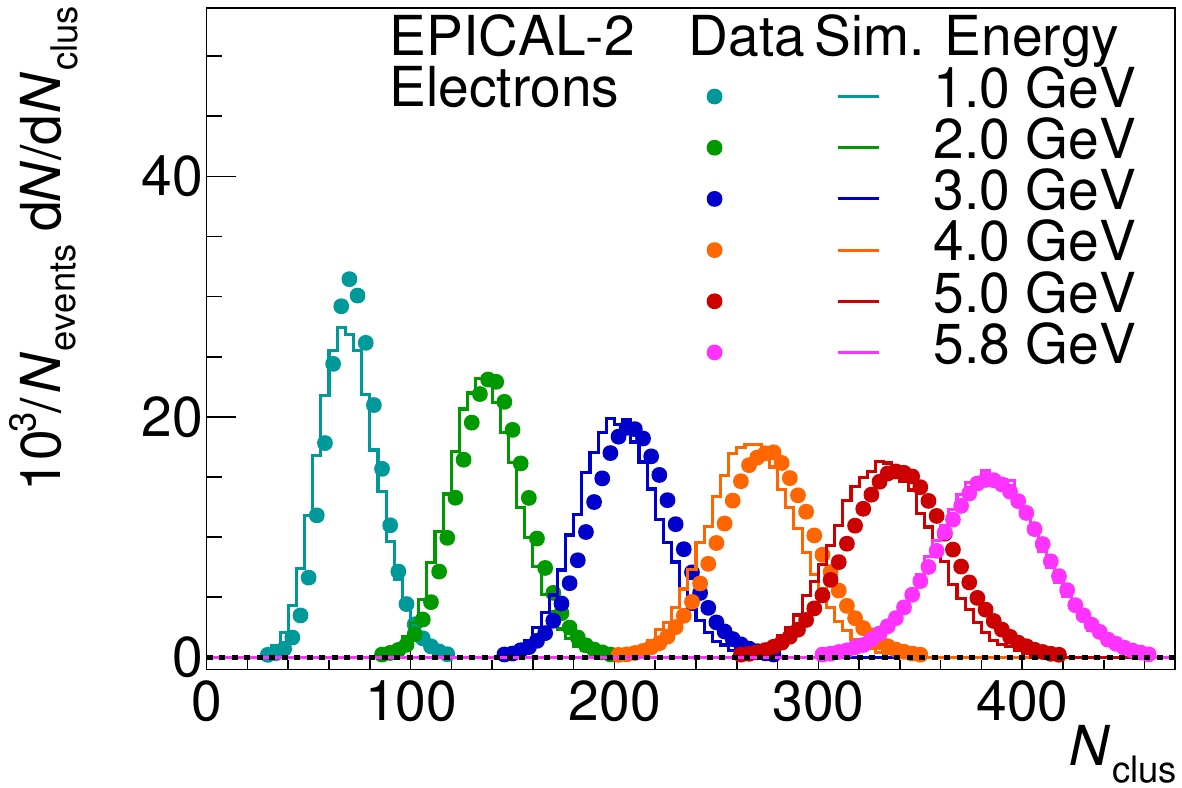}
\end{minipage}
  \caption{Comparison of data and simulation for distributions of (left) \Nhits and (right) \Nclus for all energies.}
  \label{fig:nhitsclustersdistributions}
\end{figure}

Figure~\ref{fig:nhitsclustersdistributions} shows distributions of \Nhits and  \Nclus for incident electron energies between 1.0 and 5.8\gev. The distributions show Gaussian peak structures with the most probable values increasing with energy as expected. The simulation provides a good overall description of the data, with small systematic shifts observed.

The mean $\mu$ and width $\sigma$ of these distributions are determined  arithmetically without truncation, taking the RMS for $\sigma$, and given in Tables~\ref{tab:numerical_values_mean_sigma_data} and \ref{tab:numerical_values_mean_sigma_simu} for data and simulation, respectively.
Fits to Gaussian functions, which are less sensitive to the tails of the distributions, are also performed as cross-checks.  The corresponding values for $\mu$ and particularly $\sigma$ derived from these fits are sensitive to deviations from the assumed Gaussian  form, therefore arithmetic values for $\mu$ and $\sigma$ are used.

Small differences between data and simulation are observed in \Nhits for both the highest and lowest energies considered. These are most likely to be related to the beam energy uncertainties discussed in section~\ref{sec:setup}, with slightly higher energy at low $E_0$ and lower energy at high $E_0 $ in data. Larger discrepancies are seen for \Nclus, which may be an artefact of the clustering algorithm that could be addressed with further investigation.

\begin{table}[bt]
    \caption{Mean $\mu$ and standard deviation $\sigma$ of   \Nhits and \Nclus distributions in data for electron energies 1.0--5.8\gev.}
    \centering
    \begin{tabular}{|c|r|r|r|r|}
    \hline
   \textbf{data} & \multicolumn{2}{c|}{hits} & \multicolumn{2}{c|}{clusters} \\ \hline
    $E_0$ (\gev) & \multicolumn{1}{c|}{$\mu$} & \multicolumn{1}{c|}{$\sigma$} & \multicolumn{1}{c|}{$\mu$} & \multicolumn{1}{c|}{$\sigma$} \\ \hline
  $1.0$ & $291.63  \pm 0.13$ & $71.65  \pm 0.09$ & $72.204 \pm 0.024$           & $13.467 \pm 0.017$ \\ \hline
  $2.0$ & $574.17  \pm 0.22$ & $100.05 \pm 0.16$ & $140.47 \pm 0.04\phantom{0}$ & $18.718 \pm 0.030$ \\ \hline
  $3.0$ & $862.39  \pm 0.26$ & $123.20 \pm 0.19$ & $209.33 \pm 0.05\phantom{0}$ & $23.072 \pm 0.035$ \\ \hline
  $4.0$ & $1141.96 \pm 0.25$ & $140.80 \pm 0.18$ & $275.20 \pm 0.05\phantom{0}$ & $26.099 \pm 0.033$ \\ \hline
  $5.0$ & $1417.88 \pm 0.20$ & $157.14 \pm 0.14$ & $339.04 \pm 0.04\phantom{0}$ & $28.843 \pm 0.025$ \\ \hline
  $5.8$ & $1613.99 \pm 0.29$ & $170.08 \pm 0.20$ & $383.36 \pm 0.05\phantom{0}$ & $\phantom{0}30.99 \pm 0.04\phantom{0}$ \\ \hline
    \end{tabular}
    \label{tab:numerical_values_mean_sigma_data}
\end{table}

\begin{table}[bt]
    \caption{Mean $\mu$ and the standard deviation $\sigma$ of the \Nhits  and \Nclus distributions in simulation for electron energies 1.0--5.8\gev.}
    \centering
    \begin{tabular}{|c|r|r|r|r|}
    \hline
   \textbf{simulation} & \multicolumn{2}{c|}{hits} & \multicolumn{2}{c|}{clusters} \\ \hline
    $E_0$ (\gev) & \multicolumn{1}{c|}{$\mu$} & \multicolumn{1}{c|}{$\sigma$} & \multicolumn{1}{c|}{$\mu$} & \multicolumn{1}{c|}{$\sigma$} \\ \hline
  $1.0$ & $286.20 \pm 0.32$           & $75.48 \pm 0.22$ & $69.52 \pm 0.06$ & $14.50 \pm 0.04$ \\ \hline
  $2.0$ & $568.8  \pm 0.4\phantom{0}$ & $96.71 \pm 0.30$           & $136.34 \pm 0.08$ & $17.41 \pm 0.05$ \\ \hline
  $3.0$ & $853.1  \pm 0.5\phantom{0}$ & $116.11\pm 0.35$           & $202.50 \pm 0.09$ & $20.21 \pm 0.06$ \\ \hline
  $4.0$ & $1136.7 \pm 0.6\phantom{0}$ & $131.4 \pm 0.4\phantom{0}$ & $267.93 \pm 0.10$ & $22.57 \pm 0.07$ \\ \hline
  $5.0$ & $1418.1 \pm 0.6\phantom{0}$ & $145.6 \pm 0.5\phantom{0}$ & $332.33 \pm 0.11$ & $24.66 \pm 0.08$ \\ \hline
  $5.8$ & $1643.5 \pm 0.7\phantom{0}$ & $155.7 \pm 0.5\phantom{0}$ & $383.50 \pm 0.12$ & $26.25 \pm 0.08$ \\ \hline
    \end{tabular}
    \label{tab:numerical_values_mean_sigma_simu}
\end{table}

For an ideal calorimeter, the response observable is proportional to the energy of the incident particle.
Energy-independent noise in the apparatus introduces a non-zero offset to the linear response, and this can be inferred from electron beam data or estimated from pedestal measurements.  For \epical, these are extremely low as discussed in section~\ref{sec:noise-removal}, therefore noise has a negligible effect on the detector response.
An additional complication might arise where noise is different for a detector in the presence or absence of beam\footnote{This is seen in uranium calorimeters from the induced activity.}, but as expected this is not observed in \epical.
\begin{figure}[bth]
     \begin{center}
      \includegraphics[width=0.8\textwidth]{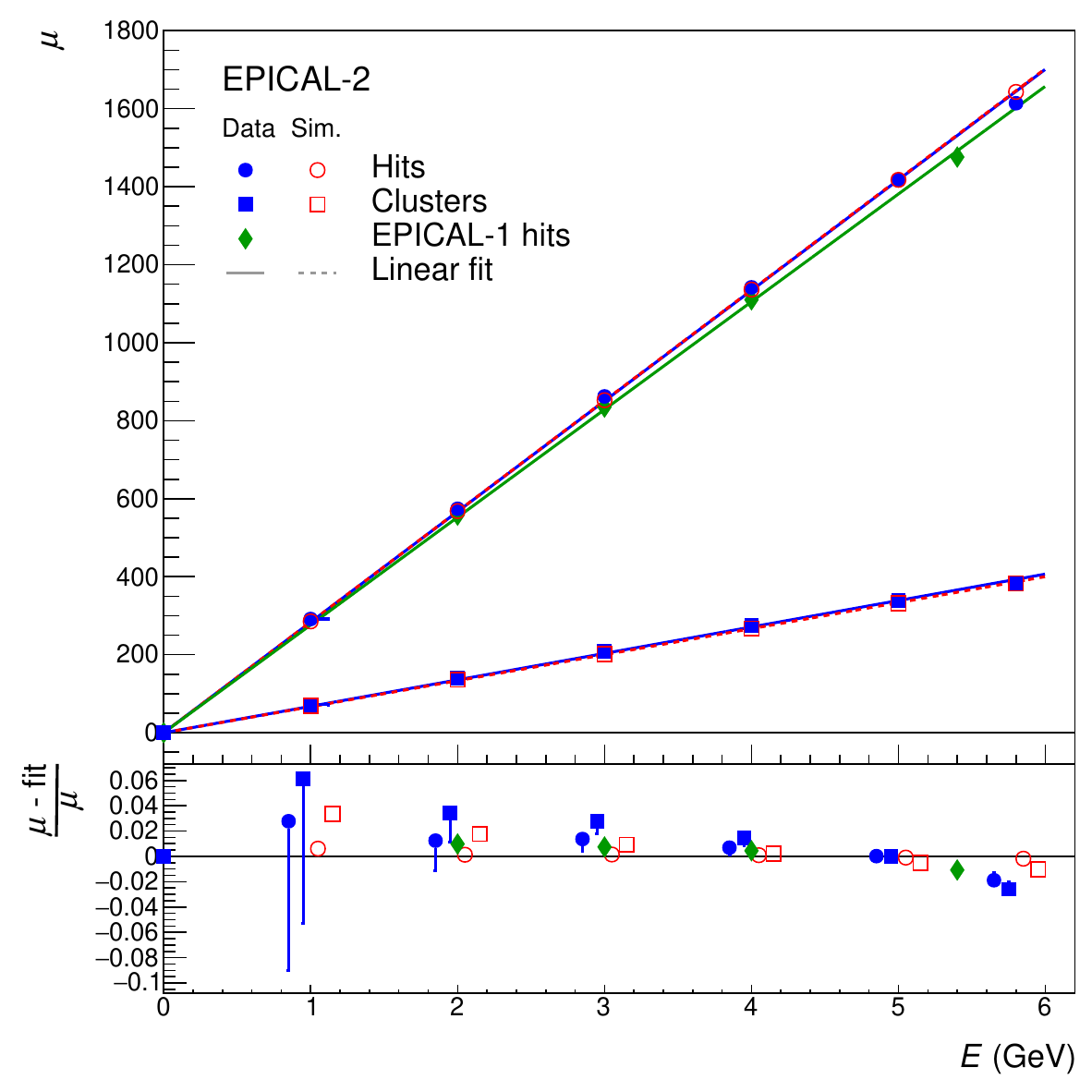}
    \end{center}
  \caption{Comparison of data and simulation as a function of electron beam energy for (upper) Energy response in terms of \Nhits and  \Nclus, including pedestal measurements at $E=0$. For  \Nhits, data from the \mimosa prototype are also  shown.
  (lower) Difference relative to the linear fit, displaced horizontally for clarity. In these ratios, a one-sided systematic uncertainty is included for the experimental data  for \epical, as discussed in the text (a corresponding uncertainty is not shown for \mimosa data).}
  \label{fig:linearity}
\end{figure}

Figure~\ref{fig:linearity} compares the mean values of the number of hits
and clusters as a function of particle energy, as well as the predictions from simulation, which provide a very good description of data.
Measurements using electron beam data and the corresponding simulations were parametrised using a linear function 
\begin{equation}
    f(E_0) = R_{\mathrm{hits,clus}} \cdot E_0 + C_{\mathrm{hits,clus}}\,.
    \label{eq:linear}
\end{equation}
This provides a very good description of simulations: for \Nhits the offset is small at $C_{\mathrm{hits}} = 2.6 \pm 0.4 = (0.009 \pm 0.001) \cdot R_{\mathrm{hits}}$, and slightly larger for \Nclus  with  $C_{\mathrm{clus}} = 4.13 \pm 0.07 = (0.063 \pm 0.001) \cdot R_{\mathrm{clus}}$. 
As discussed in section~\ref{sec:setup}, the true energy of the beam particles at DESY deviates from the nominal energy by an amount that is not fully controlled and therefore fits to data appear to show non-linearity and larger offsets.
This effect is likely to be largest at the lowest energy and so 
a fit to lower energy data alone would distort the response function and potentially yield a significant offset. This would not represent  a genuine noise-like contribution in the response.
To avoid this, pedestal measurements 
 are  included as the effective response at $E_0=0$. For the nominal measurement conditions, \ie after masking of dead or noisy pixels, the mean number of pedestal hits is small at  
$\Nhits^{\mathrm{ped}}= (2.900 \pm 0.009) \cdot 10^{-3}$. As these noise hits are dominated by isolated pixels (clusters of size 1), the same numerical value is used for $\Nclus^{\mathrm{ped}}$; the  noise contribution is negligible in practice and is included in data  at $E_0 = 0$ in figure~\ref{fig:linearity}.
For the simulations, no noise hits are found and  $C_{\mathrm{hits,clus}} \equiv 0$ in the corresponding fits.

A detailed understanding of the energy dependence requires accurate knowledge of the energy of beam particles.
As it is difficult to correct them reliably, nominal energies $E_0$ are used and a  systematic uncertainty is assigned to account for the deviations discussed above.
This leads to asymmetric systematic uncertainties on the true energies, as shown by  the error bars on the energy scale in the upper panel of figure~\ref{fig:linearity}. These  are small for most cases in this representation.
The results for each energy are fitted according equation~\ref{eq:linear},
which describes the experimental data and the simulation well. 
As expected from this procedure, all offset values $C_{\mathrm{hits,clus}}$ found are effectively zero. Conversion factors of $R_{\mathrm{hits}} =  (283.5 \pm 0.1)\, \gev^{-1} $ and $R_{\mathrm{clus}} = (67.81 \pm 0.03)\, \gev^{-1}$ are obtained for the experimental data.

The lower panel of figure~\ref{fig:linearity} shows the relative difference between  data or simulation and the corresponding fit.
The one-sided systematic uncertainty on the true energy is taken into account by an error on the energy scale in the denominator. As such, an upward error in $E_0$ corresponds to a downward error in the ratio.
In this panel, smaller deviations from linear behaviour are visible. If one were to assume just the nominal beam energies, the behaviour of the experimental data would show a deviation from linearity of the order of a few \%. In this, the number of clusters show a more visible deviation than the number of hits. Taking into account the systematic uncertainty on the true energies, the behaviour is compatible with linear. Only the highest energy data show a significant deviation and in this case  there may be an additional bias in the energy attributed that cannot be estimated  reliably (see also section~\ref{sec:setup}). 
In simulation, the  behaviour for hits is found to be linear within 1\%. For clusters, there seems to be a slightly stronger non-linearity, with the difference between clusters and hits  being very similar to that in data. It should be noted that the one-sided error on the data points due to the beam energy uncertainty is fully correlated between the results for hits and clusters at a given energy, so the difference in behaviour of clusters and hits is also significant in data.

From this, it is concluded that the response of the prototype in terms of \Nhits is consistent with very good linearity and that the deviations in the ratio are dominated by the uncertainty in the beam energy. The response in terms of \Nclus shows a small but significant non-linearity, which is seen both in data and simulations.

The response from the earlier \mimosa prototype in terms of the number of hits \cite{deHaas:2017fkf} are also included in  figure~\ref{fig:linearity}.
In the upper panel, the data from \cite{deHaas:2017fkf}  agree well with the  \epical results even on an absolute scale, which is likely to be due to similarity in  pixel size.
The similarity in behaviour of these data, despite use of very different sensor technologies, supports the conclusion that deviations from linearity are dominated by the knowledge of the DESY test beam energy rather than behaviour of the calorimeter prototype.

\subsection{Energy resolution}
\begin{figure}[tb]
   \begin{minipage}[t]{0.5\textwidth}
     \begin{center}
      \includegraphics[width=\textwidth]{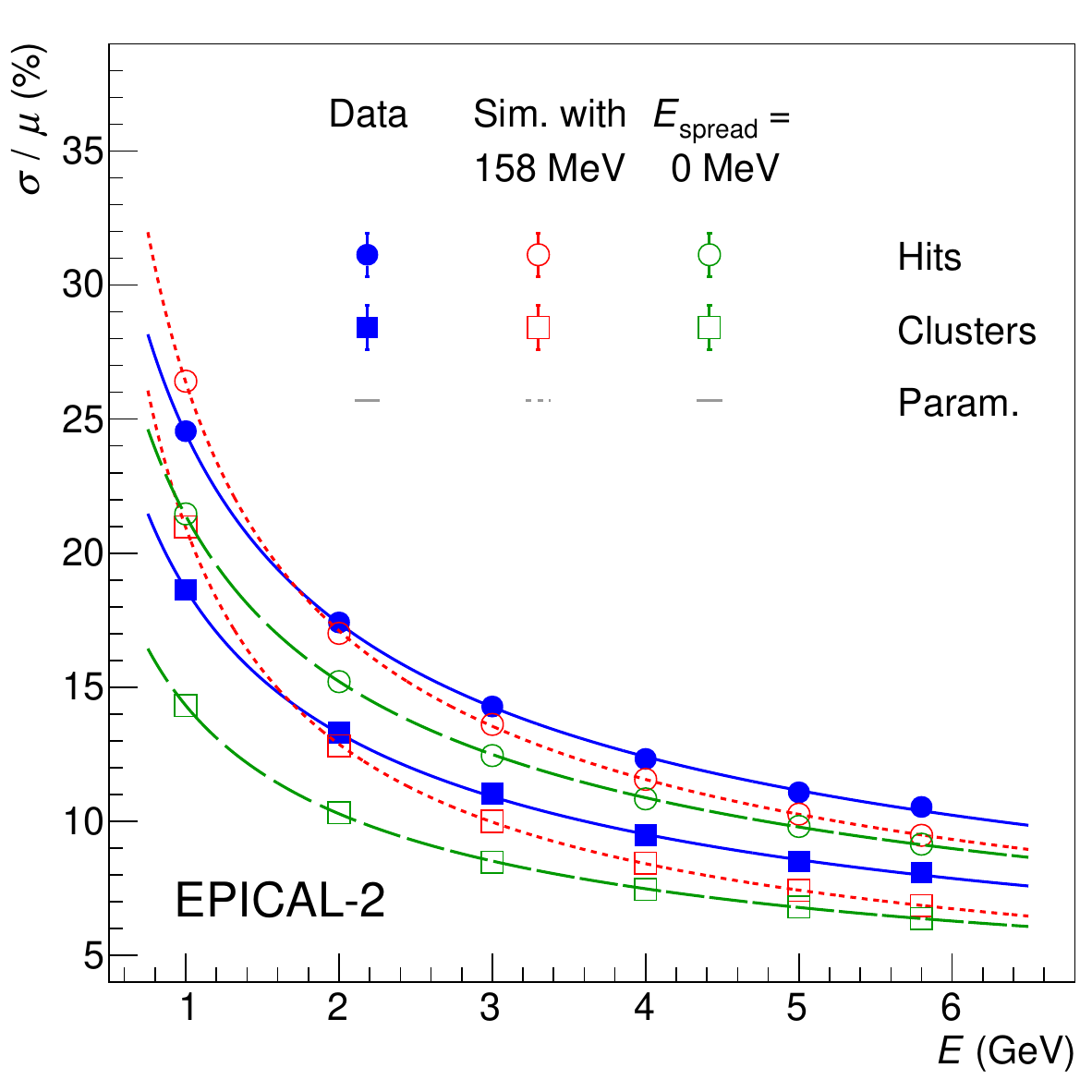}
    \end{center}
  \end{minipage}
   \begin{minipage}[t]{0.5\textwidth}
     \begin{center}
      \includegraphics[width=\textwidth]{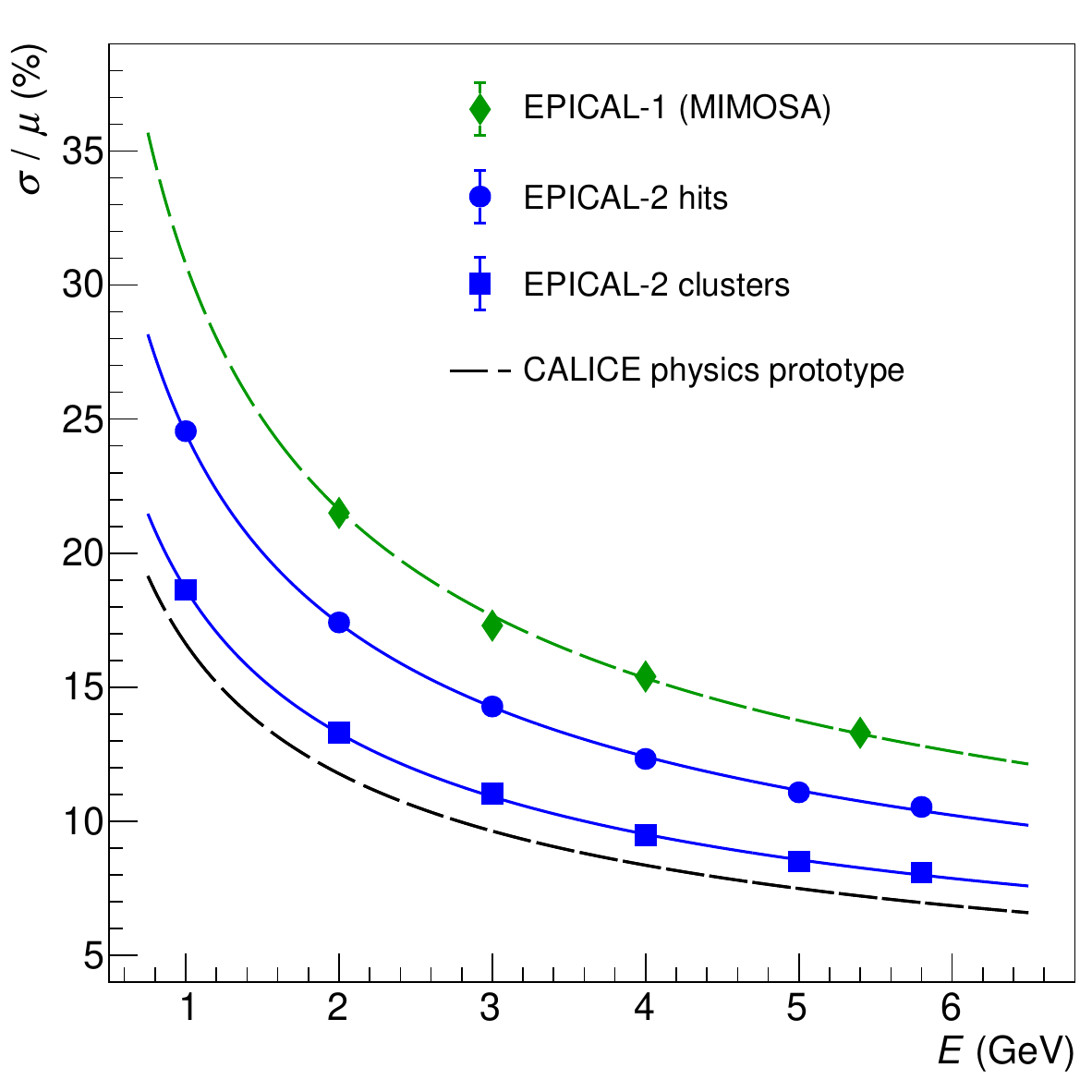}
    \end{center}
  \end{minipage}
  \caption{Energy resolution as a function of electron beam energy.
  Parametrisations are  according to equation~\ref{eq:resolution}.
  (left) Data are compared to simulations in which electron energy spreads of  $\Delta E = 158 \mev$ and $\Delta E = 0\mev$ are used.
  (right)  Comparison to data from the \mimosa prototype \cite{deHaas:2017fkf} and to the resolution of the CALICE silicon-tungsten ECAL physics prototype \cite{CALICE:2008kht}.}
  \label{fig:resolution}
\end{figure}

The energy resolution has been obtained from the distributions of figure~\ref{fig:nhitsclustersdistributions}.
The relative resolution $\sigma/\mu \equiv \sigma_E/E$ is shown as a function of beam energy in figure~\ref{fig:resolution}. Simulations have been performed assuming an energy spread of the incoming electrons of $\Delta E = 158 \mev$ as given in \cite{DESY} and follow a similar trend to data, although showing significantly better resolution than in data at high energy, and poorer at 1\gev.
To investigate the role of the energy spread, simulations have been performed without any beam energy spread. These  show smaller widths than  simulations that include finite beam energy spread, and in all cases a better resolution than in data. Neither of the two simulations model the data fully, showing that a constant energy spread does not describe the resolution as measured in \epical. As the exact contribution that the energy spread makes to the observed resolution is not known precisely but is clearly non-zero,  the measured resolution is an upper limit on the true intrinsic resolution of the detector prototype.
It is consistently observed in data and simulations that the energy resolution using \Nclus is superior to that using \Nhits.

The behaviour as a function of energy is parametrised as
\begin{equation}
\frac{\sigma_E}{E} = \frac{a}{\sqrt{E/\mathrm{GeV}}}\oplus b \oplus \frac{c}{E/\mathrm{GeV}},
\label{eq:resolution}
\end{equation}
 where $a$ and $b$ represent the contributions due to statistical fluctuations and to effects such as non-uniformity or imperfect calibration, and $c$ includes all effects independent of the beam energy including noise or a fixed beam energy spread. Table~\ref{tab:numerical_values_resolution} summarises the resolution parameters  for both data and simulation  derived from fits to hits and clusters. A good fit is obtained for data and the simulation without beam-energy spread where  parameter $c = 0$. This is not possible for the simulation including an energy spread. Here $c$ is kept as a free parameter, and the best description is obtained with values reasonably close to the value of the effective noise term that the spread should introduce ($c_{\mathrm{spread}} = 15.8$~\%).

\begin{table}[tb]
    \caption{Parameters $a$, $b$ and $c$ derived from the parametrisation of the energy resolution from hits and clusters in both data and simulation (c.f.\  figure~\ref{fig:resolution}) with the function in  equation~\ref{eq:resolution}, whereas $c$ is fixed to zero for data and the simulation scenario without beam-energy spread.}
    \centering
    \begin{tabular}{|c|c|c|c|}
    \hline 
     \textbf{hits} & \multicolumn{1}{c|}{$a \, (\%)$} & \multicolumn{1}{c|}{$b \, (\%)$} & \multicolumn{1}{c|}{$c \, (\%)$} \\ \hline
\multicolumn{1}{|c|}{data}                                          & $ 24.30 \pm 0.03 $ &	$ 2.41 \pm 0.08 $ &	-  \\ \hline
\multicolumn{1}{|c|}{sim $(E_{\text{spread}} = 0)$}                 & $ 21.27 \pm 0.06 $ &	$ 2.30 \pm 0.16 $ &	-  \\ \hline
\multicolumn{1}{|c|}{sim $(E_{\text{spread}} = 158 \, \text{MeV})$} & $ 21.58 \pm 0.25 $ &	$ 1.8 \pm 0.5 $ &	$ 15.1 \pm  0.4 $  \\ \hline \hline
     \textbf{clusters} & \multicolumn{1}{c|}{$a \, (\%)$} & \multicolumn{1}{c|}{$b \, (\%)$} & \multicolumn{1}{c|}{$c \, (\%)$} \\ \hline
\multicolumn{1}{|c|}{data}                                          & $ 18.54 \pm 0.02 $ &	$ 2.17 \pm 0.05 $ &	- \\ \hline
\multicolumn{1}{|c|}{sim $(E_{\text{spread}} = 0)$}                 & $ 14.10 \pm 0.04 $ &	$ 2.52 \pm 0.07 $ &	- \\ \hline
\multicolumn{1}{|c|}{sim $(E_{\text{spread}} = 158 \, \text{MeV})$} &	$ 14.57 \pm  0.21 $ &	$ 1.96 \pm  0.26 $ &	$ 14.93 \pm  0.23 $ \\ \hline
    \end{tabular}
    \label{tab:numerical_values_resolution}
\end{table}

Figure~\ref{fig:resolution} also compares the \epical data with measurements from two earlier  calorimeter prototypes.
For the \mimosa prototype \cite{deHaas:2017fkf}, only \Nhits has been used as response observable.
The resolution of \epical is significantly better, which is primarily  because a large fraction of the sensors were not fully operational  in \mimosa.
Data from the CALICE silicon-tungsten ECAL physics prototype \cite{CALICE:2008kht} are also presented. It  is noted that this latter result has been  determined at slightly higher energy and extrapolation to the DESY energy range may introduce additional uncertainty. Nonetheless, it is interesting to see that the resolution obtained with \Nclus in \epical is very close to that of this state-of-the-art prototype of a Si-W calorimeter with analogue readout.

\subsection{Longitudinal Shower Profile}
\begin{figure}[tb]
    \begin{minipage}[t]{0.5\textwidth}
     \begin{center}
      \includegraphics[width=\textwidth]{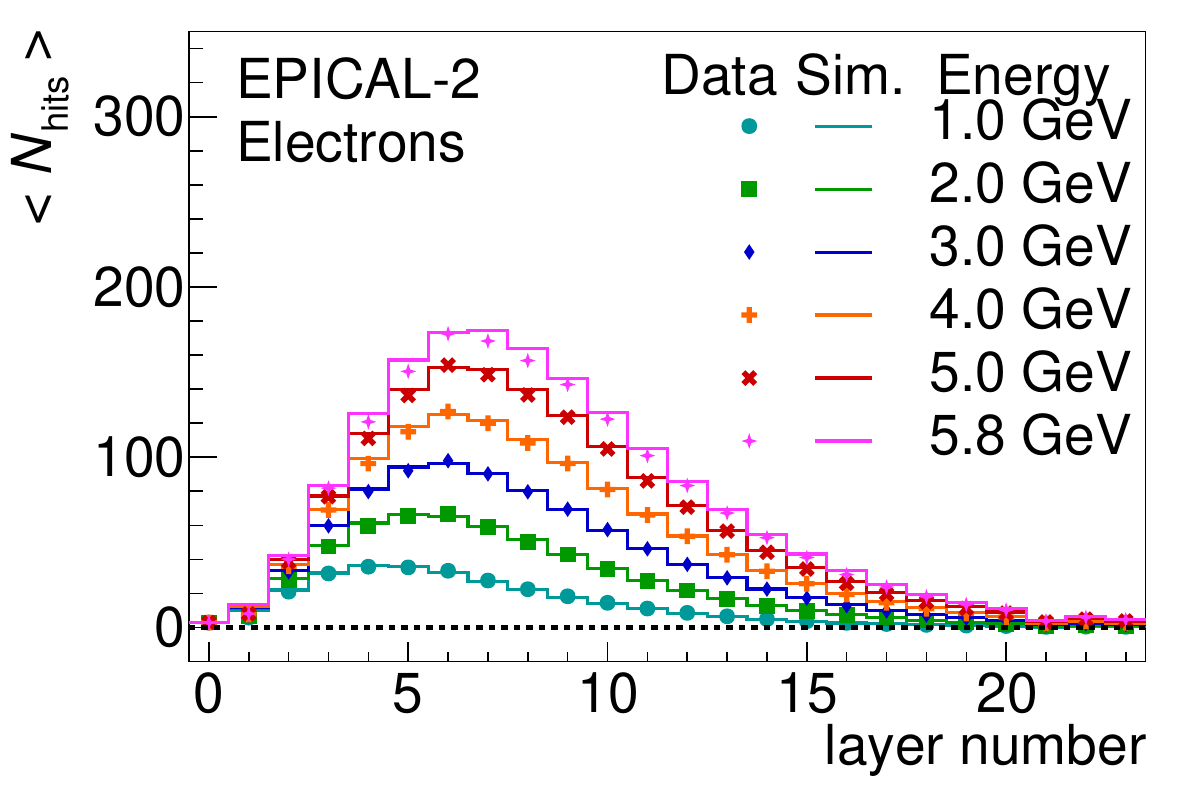}
    \end{center}
  \end{minipage}
  \begin{minipage}[t]{0.5\textwidth}
     \begin{center}
      \includegraphics[width=\textwidth]{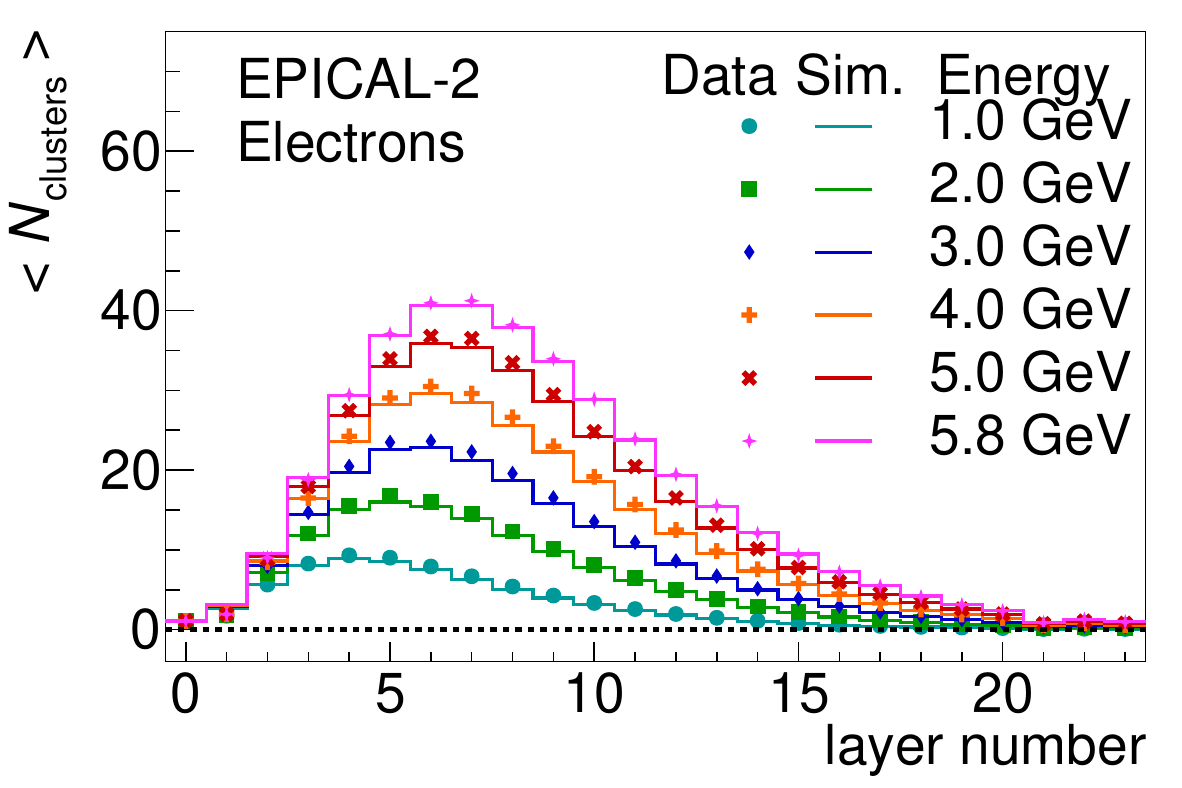}
    \end{center}
  \end{minipage}
  \caption{Longitudinal profile for the electron beam energies for the total number of (left) hits  and (right)  clusters per layer, averaged over all events. The  predictions of detailed simulations are overlaid, see text for details.}
  \label{fig:long_profile}
\end{figure}

The overall response of \epical is given by the  total number of hits (clusters) per event.
The  longitudinal profile is the average response of \epical per layer, \ie the number of hits or clusters per layer $\Nhits(l)$ ($\Nclus(l)$), where each layer is equivalent to a thickness in $z$ of $0.86X_0$,
 and reflects the evolution of the electromagnetic shower as a function of depth in material.
The measured longitudinal profiles are shown in figure~\ref{fig:long_profile}, together with the predictions from simulations.
Small discrepancies are seen: while the simulation is slightly above data for hits, it tends to underestimate the data for clusters.
Nevertheless, a good overall agreement between data and simulation is found. 
Also the reduction of the number of hits and clusters in layer 21, as a result of the single inactive ALPIDE chip in this layer during test-beam data taking, is correctly modelled by simulation.
The depth at which the maximum intensity of the shower is observed increases with electron energy as expected, for both simulation and data.\footnote{The data are compared directly to simulation rather than fitted to a gamma distribution \cite{Longo:1975wb}, which is itself a parametrisation of simulated data.}  

\subsection{Lateral Shower Profile}
\label{sec:lateral}
The high granularity of the \epical prototype allows the lateral profile of energy deposits within the electromagnetic shower to be determined with considerable precision.  Here, the emphasis is on the average behaviour of the lateral profile.

Several methods for calculating a 3d vector representing the shower axis were considered for this analysis, including 3d fitting of the hits in the event and the  method used for \mimosa correlating clusters in layers~3/4 with clusters in layer~0 \cite{Zhang:2317153}. 
In the following analysis, the shower axis is  defined as a vector perpendicular to the $x$-$y$ plane, with transverse coordinates defined by the mean position of hits in layer~0 (after alignment corrections). This provides good agreement with hits in the more upstream layers that are a reliable estimator of the trajectory of the incident particle, before significant development of the shower has taken place.

The distance between the centre of each pixel and the determined shower axis in the same layer is used to form the hit density within each of several annular regions.
A geometric calculation is used to account for the insensitive area between the two chips in a layer. 
This profile is evaluated for several illustrative layers in shower development: layer~2 represents early shower development, layer~5 represents an approximate shower maximum, and layers~8 and 11 represent the early and late tail regions, respectively. A profile of the longitudinal hit density is also produced for a range of annular regions.
\begin{figure}[t]
    \begin{minipage}[t]{0.5\textwidth}
     \begin{center}
      \includegraphics[width=\textwidth]{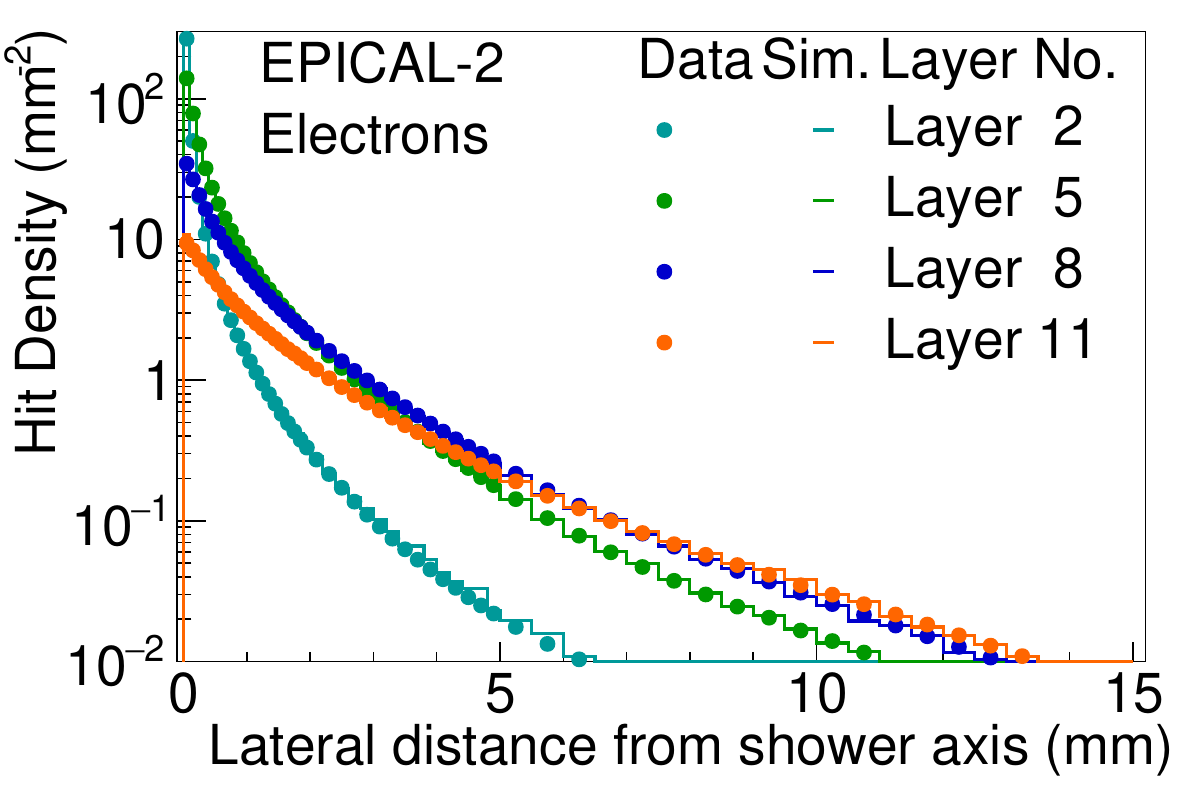}
    \end{center}
  \end{minipage}
  \begin{minipage}[t]{0.5\textwidth}
     \begin{center}
      \includegraphics[width=\textwidth]{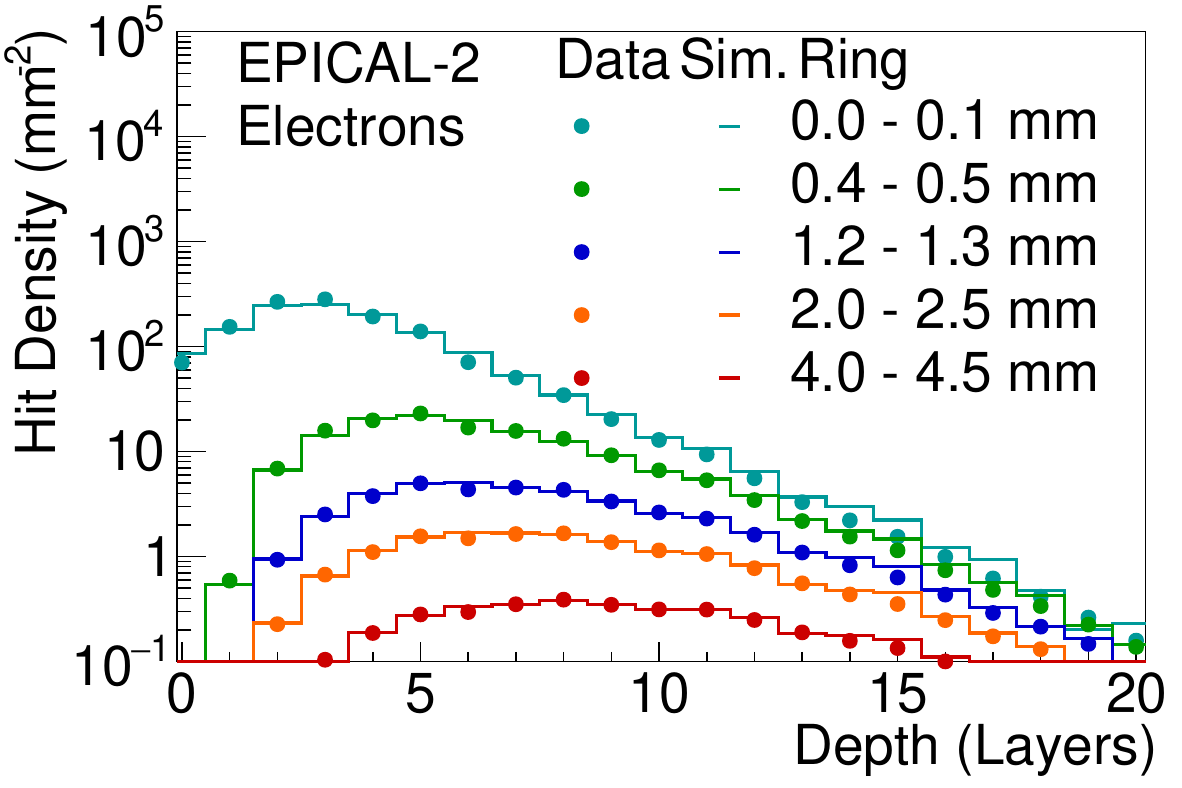}
    \end{center}
  \end{minipage}
  \caption{Comparison of 5\gev electron data with simulation, (left) lateral profile for a range of layers and (right) longitudinal hit density profile, for a range of annular regions around the shower axis.}
  \label{fig:lateral_profile}
\end{figure}

From the results shown in figure~\ref{fig:lateral_profile}, there is little evidence of  saturation of hits. The hit density plateaus significantly as the shower develops and spreads out. For very small radii the hit density is larger in earlier layers, the largest number of hits at the shower maximum (layer~5) is the product of the larger integral under the hit density curve. The longitudinal hit density profile shows that the maximum hit density for each annular region moves to larger depths for increasing radius such that by layer~11 the density of hits in the inner two rings become comparable.
As a consequence of this shower evolution, the shower maximum is at $\approx$ layer~5, while the greatest hit density occurs in layer~3, at $\approx$ 300~hits/mm$^2$. 
This is significantly less than complete saturation, which occurs at around  1272~hits/mm$^2$ for this prototype.
Saturation effects may still play a role because for individual showers,  fluctuations may cause the density to approach this limit. However, the average value of the density is far below this saturation limit, and any effects from local shower fluctuations should affect only a very small fraction of the total shower distribution, so that saturation is expected to be negligible for the total number of hits at these energies.

\section{Conclusion}
\label{sec:conclusion}

The fully digital electromagnetic calorimeter prototype 
\epical has been constructed using a Si-W sandwich structure with 24 sampling layers, each of which consists of two ALPIDE pixel sensors. The device has an active cross section of
\SI[product-units=power]{27.6 x 29.9}{\mm} and a thickness of 84~mm, corresponding to almost 20 radiation lengths. 
With the help of advanced ultra-thin connection technology, the prototype constructed is very compact, with an effective Moli\`ere radius that is expected to be close to that of tungsten.
Measurements with cosmic muons have been used to perform a relative alignment of the sensors with an accuracy of better than 5~$\mathrm{\mu m}$ and to account for the different effective thresholds of each sensor. The sensors show extremely low noise, achieving a 
fake hit rate (noise) below $10^{-10}$/pixel/event after masking 0.015~\% of the pixels. 

Measurements with electrons and positrons with energies between 1.0 and 5.8\gev have been performed at the DESY~II Test Beam Facility.
The ALPIDE sensors used, which have been developed for tracking applications, are shown to be fully capable of measurements of electromagnetic showers with their high instantaneous particle densities. 
Detailed simulations using \Allpixtwo describe the behaviour of the detector very well, for both  the size and shape of individual clusters and the distribution of more global variables.

The number of hits \Nhits and the number of clusters \Nclus have been used as observables for the energy response of the calorimeter. Both show a close to linear dependence on the beam energy. Small deviations from linearity, which are not fully reproduced in simulations, are attributed to uncertainties in the known particle energy. The relative energy resolution has also been obtained from these two observables, and in both cases shows the expected dependence on the beam energy. 

Interestingly, \Nhits shows a slightly better linearity than  \Nclus, while \Nclus provides a  significantly better energy resolution than does \Nhits.
These differences between \Nhits and \Nclus are described well by simulations.
The better resolution for \Nclus is associated with lower fluctuations in the charge cloud from single charged particles, while the initial clustering algorithm used  makes \Nclus  more susceptible to merging of close-by clusters, leading to stronger non-linearity.
Development of alternative algorithms for clustering of hits will be investigated in the future using data from a wider range of energies. In addition, the problem could be partially mitigated by the use of a non-zero back bias voltage, but this is beyond the scope of this paper.

The energy resolution obtained from \Nhits is already significantly better than the one of the previous digital pixel prototype \mimosa \cite{deHaas:2017fkf}. Using the higher resolution observable \Nclus  and without correction for the intrinsic momentum spread of the test beam, we obtain:
\begin{equation}
\frac{\sigma_E}{E} = \frac{18.5\,\%}{\sqrt{E/\mathrm{GeV}}}\oplus 2.2\,\%,
\label{eq:resolution2}
\end{equation}
which is very close to results for a state-of-the-art analogue Si-W calorimeter \cite{CALICE:2008kht}.
With its very good performance, the technology used here certainly qualifies for the pixel layers in the ALICE FoCal detector \cite{Focal-loi}. 
While the current analysis with \epical uses only relatively low-energy particles, the performance of  \mimosa at higher energies is also reasonably good \cite{deHaas:2017fkf}. This has to be confirmed by equivalent measurements with \epical, but together these measurements already demonstrate the potential of digital ECAL technology for future applications in high-energy physics.
Optimisation of the CMOS sensor design beyond the currently used tracking chip for the application in a calorimeter will most certainly further improve the performance of this technology.

First exploratory studies of the shape of electromagnetic showers have also been performed. Longitudinal and lateral profiles have been obtained from the measurement and agree well with simulations, reproducing the expected global features of the evolution of an electromagnetic shower.
The small differences observed between data and simulation, as well as further  differential studies exploiting the unprecedented detail  accessible with the highly granular  \epical prototype, including an event-by-event variation of shapes, will be investigated in a future publication.

\acknowledgments
We would like to thank J.A.~Hasenbichler for providing the results of the TCAD simulations, J.~Schambach for the multi-channel transition board design, and M.~Bonora and M.~Lupi for help in installation/implementation and adaptation of FPGA firmware and software. We also thank the DESY test beam coordinators and the supporting people at DESY for the usage of the test beam, and M.~Stanitzki for useful discussions.

\bibliography{Main}

\end{document}